%% file: main.tex
\documentclass[times,onecolumn]{aastex62}

\usepackage{cases}
\usepackage{graphicx}
\usepackage{subfigure}
\usepackage{natbib}
\usepackage{color}
\usepackage{tabularx}
\usepackage{bm}
\usepackage{threeparttable}

\newcommand{\thetae}{\theta_{\rm E}}
\newcommand{\pie}{\pi_{\rm E}}

\newcommand{\te}{t_{\rm E}}
\newcommand{\event}{KMT-2016-BLG-1836}

\newcommand{\Sp}{{\it Spitzer}}
\newcommand{\hjd}{${\rm HJD}^{\prime}$}

\DeclareGraphicsExtensions{.pdf,.png,.jpg}

\shorttitle{}
\shortauthors{Yang et al.}

\begin{document}
\title{KMT-2016-BLG-1836Lb: A Super-Jovian Planet From A High-Cadence Microlensing Field}

\correspondingauthor{Weicheng Zang}
\email{zangwc17@mails.tsinghua.edu.cn}

\author[0000-0003-0626-8465]{Hongjing Yang}
\affiliation{Department of Astronomy and Tsinghua Centre for Astrophysics, Tsinghua University, Beijing 100084, China}
\affiliation{Department of Astronomy, Xiamen University, Xiamen 361005, China}

\author{Xiangyu Zhang}
\affiliation{Department of Astronomy and Tsinghua Centre for Astrophysics, Tsinghua University, Beijing 100084, China}

\author{Kyu-Ha Hwang}
\affiliation{Korea Astronomy and Space Science Institute, Daejon 34055, Republic of Korea}

\author[0000-0001-6000-3463]{Weicheng Zang}
\affiliation{Department of Astronomy and Tsinghua Centre for Astrophysics, Tsinghua University, Beijing 100084, China}

\author{Andrew Gould}
\affiliation{Max-Planck-Institute for Astronomy, K\"onigstuhl 17, 69117 Heidelberg, Germany}
\affiliation{Department of Astronomy, Ohio State University, 140 W. 18th Ave., Columbus, OH 43210, USA}

\author{Tianshu Wang}
\affiliation{Department of Astronomy and Tsinghua Centre for Astrophysics, Tsinghua University, Beijing 100084, China}

\author{Shude Mao}
\affiliation{Department of Astronomy and Tsinghua Centre for Astrophysics, Tsinghua University, Beijing 100084, China}
\affiliation{National Astronomical Observatories, Chinese Academy of Sciences, Beijing 100101, China}


\author{Michael D. Albrow}
\affiliation{University of Canterbury, Department of Physics and Astronomy, Private Bag 4800, Christchurch 8020, New Zealand}

\author{Sun-Ju Chung}
\affiliation{Korea Astronomy and Space Science Institute, Daejon 34055, Republic of Korea}
\affiliation{University of Science and Technology, Korea, (UST), 217 Gajeong-ro Yuseong-gu, Daejeon 34113, Republic of Korea}

\author{Cheongho Han}
\affiliation{Department of Physics, Chungbuk National University, Cheongju 28644, Republic of Korea}

\author{Youn Kil Jung}
\affiliation{Korea Astronomy and Space Science Institute, Daejon 34055, Republic of Korea}

\author{Yoon-Hyun Ryu}
\affiliation{Korea Astronomy and Space Science Institute, Daejon 34055, Republic of Korea}

\author{In-Gu Shin}
\affiliation{Korea Astronomy and Space Science Institute, Daejon 34055, Republic of Korea}

\author{Yossi Shvartzvald}
\affiliation{IPAC, Mail Code 100-22, Caltech, 1200 E. California Blvd., Pasadena, CA 91125, USA}

\author{Jennifer~C.~Yee}
\affiliation{Center for Astrophysics $|$ Havard \& Smithsonian, 60 Garden St.,Cambridge, MA 02138, USA}

\author{Wei Zhu}
\affiliation{Canadian Institute for Theoretical Astrophysics, University of Toronto, 60 St George Street, Toronto, ON M5S 3H8, Canada}

\author[0000-0001-7506-5640]{Matthew T. Penny}
\affiliation{Department of Astronomy, The Ohio State University, 140 W. 18th Avenue, Columbus, OH 43210, USA}

\author{Pascal Fouqu\'e}
\affiliation{CFHT Corporation, 65-1238 Mamalahoa Hwy, Kamuela, Hawaii 96743, USA}
\affiliation{Universit\'e de Toulouse, UPS-OMP, IRAP, Toulouse, France}

\author{Sang-Mok Cha}
\affiliation{Korea Astronomy and Space Science Institute, Daejon 34055, Republic of Korea}
\affiliation{School of Space Research, Kyung Hee University, Yongin, Kyeonggi 17104, Republic of Korea} 

\author{Dong-Jin Kim}
\affiliation{Korea Astronomy and Space Science Institute, Daejon 34055, Republic of Korea}

\author{Hyoun-Woo Kim}
\affiliation{Korea Astronomy and Space Science Institute, Daejon 34055, Republic of Korea}

\author{Seung-Lee Kim}
\affiliation{Korea Astronomy and Space Science Institute, Daejon 34055, Republic of Korea}
\affiliation{University of Science and Technology, Korea, (UST), 217 Gajeong-ro Yuseong-gu, Daejeon 34113, Republic of Korea}

\author{Chung-Uk Lee}
\affiliation{Korea Astronomy and Space Science Institute, Daejon 34055, Republic of Korea}
\affiliation{University of Science and Technology, Korea, (UST), 217 Gajeong-ro Yuseong-gu, Daejeon 34113, Republic of Korea}

\author{Dong-Joo Lee}
\affiliation{Korea Astronomy and Space Science Institute, Daejon 34055, Republic of Korea}

\author{Yongseok Lee}
\affiliation{Korea Astronomy and Space Science Institute, Daejon 34055, Republic of Korea}
\affiliation{School of Space Research, Kyung Hee University, Yongin, Kyeonggi 17104, Republic of Korea}

\author{Byeong-Gon Park}
\affiliation{Korea Astronomy and Space Science Institute, Daejon 34055, Republic of Korea}
\affiliation{Korea University of Science and Technology, 217 Gajeong-ro, Yuseong-gu, Daejeon 34113, Republic of Korea}

\author{Richard W. Pogge}
\affiliation{Department of Astronomy, Ohio State University, 140 W. 18th Ave., Columbus, OH 43210, USA}

\input{abstract}

\section{Introduction}\label{intro}
\input{intro}

\section{Observations}\label{obser}

\input{obser}

\section{Light curve analysis}\label{model}
\input{model}

\section{Source Properties}\label{source}
\input{source}

\section{Lens Properties}\label{lens}
\input{lens}

\section{Discussion}\label{dis}
\input{dis}

\bibliography{Zang.bib}

\input{table.tex}

\input{figure.tex}

\end{document}

%% file: abstract.tex
\begin{abstract}
We report the discovery of a super-Jovian planet in the microlensing event KMT-2016-BLG-1836, which was found by the Korea Microlensing Telescope Network's high-cadence observations ($\Gamma \sim 4~{\rm hr}^{-1}$). The planet-host mass ratio $q \sim 0.004$. A Bayesian analysis indicates that the planetary system is composed of a super-Jovian $M_{\rm planet} = 2.2_{-1.1}^{+1.9} M_{J}$ planet orbiting an M or K dwarf $M_{\rm host} = 0.49_{-0.25}^{+0.38} M_{\odot}$, at a distance of $D_{\rm L} = 7.1_{-2.4}^{+0.8}$~kpc. The projected planet-host separation is $3.5^{+1.1}_{-0.9}$~AU, implying that the planet is located beyond the snowline of the host star. Future high-resolution images can potentially strongly constrain the lens brightness and thus the mass and distance of the planetary system. Without considering detailed detection efficiency, selection or publication biases, we find a potential ``mass ratio desert'' at $-3.7 \lesssim \log q \lesssim -3.0$ for the 31 published KMTNet planets.
\end{abstract}


%% file: intro.tex
Since the first robust detection of a microlens planet in 2003 \citep{OB03235}, more than 70\footnote{\url{http://exoplanetarchive.ipac.caltech.edu} as of 2019 July 17} extrasolar planets have been detected by the microlensing method \citep{Shude1991,Andy1992}. Unlike other methods that rely on the light from the host stars, the microlensing method uses the light from a background source deflected by the gravitational potential of an aligned foreground planetary system. Thus, microlensing can detect planets around all types of stellar objects at various Galactocentric distances \citep[e.g.,][]{Novati2015,Zhu2017spitzer}. 

The typical Einstein timescale $\te$ for microlensing events is about $20$~days, and the half-duration of a planetary perturbation \citep{Andy1992} is
\begin{equation}\label{tp}
   t_{p} \sim \te\sqrt{q} \to 5(q/10^{-4})^{1/2}{\rm hr}, 
\end{equation}
where $q$ is the planet-host mass ratio. Assuming that about 10 data points are needed to cover the planetary perturbation, a cadence of $\Gamma \sim 1~{\rm hr}^{-1}$ would be required to discover ``Neptunes'' and $\Gamma \sim 4~{\rm hr}^{-1}$ would be required to detect Earths \citep{Henderson2014}. In addition, because the optical depth to microlensing toward the Galactic bulge is only $\tau \sim 10^{-6}$ \citep{Sumidepth,OGLE4optical}, a large area (10--100 ${\rm deg}^2$) must be monitored to find a large number of microlensing events and thus planetary events. 

For many years, most microlensing planets were discovered by a combination of wide-area surveys for finding microlensing events and intensive follow-up observations for capturing the planetary perturbation \citep{Andy1992}. This strategy mainly focused on high-magnification events \citep[e.g.,][]{OB050071} which intrinsically have high sensitivity to planets \citep{Griest1998}. Another strategy to find microlensing planets is to conduct wide-area, high-cadence surveys toward the Galactic bulge. The Korea Microlensing Telescope Network (KMTNet, \citealt{KMT2016}), continuously monitors a broad area at relatively high-cadence toward the Galactic bulge from three 1.6~m telescopes equipped with 4 ${\rm deg}^2$ FOV cameras at the Cerro Tololo Inter-American Observatory (CTIO) in Chile (KMTC), the South African Astronomical Observatory (SAAO) in South Africa (KMTS), and the Siding Spring Observatory (SSO) in Australia (KMTA). It aims to simultaneously find microlensing events and characterize the planetary perturbation without the need for follow-up observations. 

In its 2015 commissioning season, KMTNet followed this strategy and observed four fields at a very high cadence of $\Gamma = 6~{\rm hr}^{-1}$. Beginning in 2016, KMTNet monitors a total of (3, 7, 11, 2) fields at cadences of $\Gamma \sim (4, 1, 0.4, 0.2)~ {\rm hr}^{-1}$. See Figure 12 of \cite{Kim2018a}. This new strategy mainly aims to support \Sp\ microlensing campaign \citep{GouldSp1,GouldSp2,GouldSp4,GouldSp3,GouldSp5,GouldSp6} and find more planets over a much broader area. So far, this new strategy has detected \textbf{30} planets in 2016--2018\footnote{OGLE-2016-BLG-0263Lb \citep{OB160263}, OGLE-2016-BLG-0596Lb \citep{OB160596}, OGLE-2016-BLG-0613Lb \citep{OB160613}, OGLE-2016-BLG-1067Lb \citep{OB161067}, OGLE-2016-BLG-1190Lb \citep{OB161190}, OGLE-2016-BLG-1195Lb \citep{OB161195}, OGLE-2016-BLG-1227Lb \citep{OB161227}, KMT-2016-BLG-0212Lb \citep{KB160212}, KMT-2016-BLG-1107Lb \citep{KB161107}, KMT-2016-BLG-1397Lb \citep{KB161397}, KMT-2016-BLG-1820Lb \citep{KB161820}, MOA-2016-BLG-319Lb \citep{MB16319}, OGLE-2017-BLG-0173Lb \citep{OB170173}, OGLE-2017-BLG-0373Lb \citep{OB170373}, OGLE-2017-BLG-0482Lb \citep{OB170482}, OGLE-2017-BLG-1140Lb \citep{OB171140}, OGLE-2017-BLG-1434Lb \citep{OB171434}, OGLE-2017-BLG-1522Lb \citep{OB171522}, KMT-2017-BLG-0165Lb \citep{KB170165}, KMT-2017-BLG-1038Lb \citep{KB171038}, KMT-2017-BLG-1146Lb \citep{KB171038}, OGLE-2018-BLG-0532Lb \citep{OB180532}, OGLE-2018-BLG-0596Lb \citep{OB180596}, OGLE-2018-BLG-0740Lb \citep{OB180740}, OGLE-2018-BLG-1011Lbc \citep{OB181011}, OGLE-2018-BLG-1700Lb \citep{OB181700}, KMT-2018-BLG-0029Lb \citep{KB180029}, KMT-2018-BLG-1292Lb \citep{KB181292}, and KMT-2018-BLG-1990Lb \citep{KB181990}.}, including an Earth-mass planet found by a cadence of $\Gamma \sim 4~{\rm hr}^{-1}$ \citep{OB161195}, and a super-Jovian planet found by a cadence of $\Gamma \sim 0.2 {\rm hr}^{-1}$ \citep{KB181292}.

Here we report the analysis of a super-Jovian planet KMT-2016-BLG-1836Lb, which was detected by KMTNet's $\Gamma \sim 4~{\rm hr}^{-1}$ observations.  The paper is structured as follows. In Section \ref{obser}, we introduce the KMTNet observations of this event. We then describe the light curve modeling process in Section \ref{model}, the properties of the microlens source in Section \ref{source}, and the physical parameters of the planetary system in Section \ref{lens}. Finally, we discuss the mass ratio distributions of 31 published KMTNet planets in Section \ref{dis}. 




%% file: obser.tex
\event\ was at equatorial coordinates $(\alpha, \delta)_{\rm J2000}$ = (17:53:00.08, $-30$:02:26.70), corresponding to Galactic coordinates $(\ell,b)=(-0.12,-1.95)$. It was found by applying the KMTNet event-finding algorithm \citep{Kim2018a} to the 2016 KMTNet survey data \citep{kmt2016data}, and the apparently amplified flux of a KMTNet catalog-star $I = 19.20 \pm 0.13$ derived from the OGLE-III star catalog \citep{OGLEIII} led to the detection of this microlensing event. \event\ was located in two slightly offset fields BLG02 and BLG42, with a nominal combined cadence of $\Gamma = 4~{\rm hr}^{-1}$. In fact, the cadence of KMTA and KMTS was altered to $\Gamma = 6~{\rm hr}^{-1}$ from April 23 to June 16 ($7501 < {\rm HJD}^{\prime} < 7555, {\rm HJD}^{\prime} = {\rm HJD} - 2450000$) to support the \emph{Kepler} ${\it K2}$C9 campaign \citep{GouldK2,K2C9,kmtk2}. This higher cadence block came toward the end of the event and after the planetary perturbation. The majority of observations were taken in the $I$-band, with about $10\%$ of the KMTC images and $5\%$ of the KMTS images taken in the $V$-band for the color measurement of microlens sources. All data for the light curve analyses were reduced using the pySIS software package \citep{pysis}, a variant of difference image analysis \citep{Alard1998}. For the source color measurement and the color-magnitude diagram (CMD), we additionally conduct pyDIA photometry\footnote{MichaelDAlbrow/pyDIA: Initial Release on Github, doi:10.5281/zenodo.268049} for the KMTC02 data, which simultaneously yields field-star photometry on the same system as the light curve.

%% file: model.tex
Figure \ref{lc} shows the \event\ data together with the best-fit model. The light curve shows a bump (\hjd~$\sim 7493$) after the peak of an otherwise normal \cite{Paczynski1986} point-lens light curve. The bump could be a binary-lensing (2L1S) anomaly that is generally produced by caustic-crossing \citep[e.g.,][]{OB150966} or cusp approach \citep[e.g.,][]{OB161195} of the lensed star, or the second peak of a binary-source event (1L2S), which is the superposition of two point-lens events generated by two source stars \citep{Gaudi1998, Han_BS}. Thus, we perform both binary-lens and binary-source analyses in this section. 

\subsection{Binary-lens (2L1S) Modeling}\label{BL}
A standard binary lens model has seven parameters to calculate the magnification, $A(t)$. Three ($t_0$, $u_0$, $\te$)  of these parameters describe a point-lens event \citep{Paczynski1986}: the time of the maximum magnification, the minimum impact parameter in units of the angular Einstein radius $\thetae$, and the Einstein radius crossing time. The next three ($q$, $s$, $\alpha$) define the binary geometry: the binary mass ratio, the projected separation between the binary components normalized to the Einstein radius, and the angle between the source trajectory and the binary axis in the lens plane. The last parameter is the source radius normalized by the Einstein radius, $\rho = \theta_*/\theta_{\rm E}$. In addition, for each data set $i$, two flux parameters ($f_{{\rm S},i}$, $f_{{\rm B},i}$) represent the flux of the source star and the blend flux. The observed flux, $f_{i}(t)$, calculated from the model is

\begin{equation}
    f_{i}(t) = f_{{\rm S},i} A(t) + f_{{\rm B},i}.
\end{equation}

We locate the $\chi^2$ minima by a searching over a grid of parameters ($\log s, \log q, \alpha$). The grids consist of 21 values equally spaced between $-1\leq\log s \leq1$, 10 values equally spaced between $0^{\circ}\leq \alpha < 360^{\circ}$, and 51 values equally spaced between $-5\leq \log q \leq0$. For each set of ($\log s, \log q, \alpha$), we fix $\log q$, $\log s$, $\rho=0.001$, and free $t_0, u_0, \te, \alpha$. We find the minimum $\chi^2$ by Markov chain Monte Carlo (MCMC) $\chi^2$ minimization using the \texttt{emcee} ensemble sampler \citep{emcee}. The upper panel of Figure \ref{grid} shows the $\chi^2$ distribution in the ($\log s, \log q$) plane from the grid search, which indicates the distinct minima are within $-0.3\leq\log s \leq0.3$ and $-5\leq\log q \leq-1$. We therefore conduct a denser grid search, which consists of 61 values equally spaced between $-0.3\leq\log s \leq0.3$, 10 values equally spaced between $0^{\circ}\leq \alpha < 360^{\circ}$, and 41 values equally spaced between $-5\leq \log q \leq-1$. As a result, we find four distinct minima and label them as ``A'', ``B'', ``C'' and ``D'' in the lower panel of Figure \ref{grid}. We then investigate the best-fit model with all free parameters. Table \ref{parm1} shows best-fit parameters of the four solutions from MCMC. The MCMC results show that the solution ``B'' is the best-fit model, while the solution ``A'' is disfavored by $\Delta\chi^2 \sim 16$. We note that these two solutions are related by the so-called close-wide degeneracy and approximately take $s \leftrightarrow s^{-1}$ \citep{Griest1998, Dominik1999}, so we label them by ``Close'' (solution B, $s < 1$) and ``Wide'' (solution A, $s > 1$) in the following analysis. The solutions ``C'' and ``D'' are disfavored by $\Delta\chi^2 \sim 474$ and $\Delta\chi^2 \sim 235$, respectively, so we exclude these two solutions. For both the solutions ``Close'' and ``Wide'', the data are consistent with a point-source model within $\sim 2\sigma$ level, and the upper limit for $\rho$ is $2.0 \times 10^{-3}$ for the solution ``Close'' and $2.8 \times 10^{-3}$ for the solution ``Wide''. The best-fit model curves for the two solutions are shown in Figure \ref{lc}, and their magnification maps are shown in Figure \ref{cau}.


In addition, we check whether the fit further improves by considering the microlens-parallax effect,
\begin{equation}
\bm{\pi}_{\rm E} = \frac{\pi_{\rm rel}}{\thetae}\frac{\bm{\mu}_{\rm rel}}{\mu_{\rm rel}}~,   
\end{equation}
where $(\pi_{\rm rel}, \bm{\mu}_{\rm rel})$ are the lens-source relative (parallax, proper motion), which is caused by the orbital acceleration of Earth \citep{Gould1992}. We also fit $u_0 > 0$ and $u_0 < 0$ solutions to consider the ``ecliptic degeneracy'' \citep{OB09020}. To facilitate the further discussion of these solutions, we label them by $C_\pm$ or $W_\pm$. The letter stands for ``Close'' ($s < 1$) or ``Wide'' ($s > 1$), while the subscript refers to the sign of $u_0$. The addition of parallax to the model does not significantly improve the fit, providing an improvement of $\Delta \chi^2 < 3.0$ for the $C_\pm$ solutions and $\Delta \chi^2 < 1.7$ for the $W_\pm$ solutions. However, we find that the east component of the parallax vector $\pi_{\rm E, E}$ is well constrained for all the solutions, while the constraint on the north component $\pi_{\rm E, N}$ is considerably weaker. Table \ref{parm2} shows best-fit parameters of the standard binary-lens model, $C_\pm$ and $W_\pm$ solutions, and Figure \ref{pie} shows the likelihood distribution of $(\pi_{\rm E,N}, \pi_{\rm E,E})$ from MCMC. 


\subsection{Binary-source (1L2S) Modeling}\label{BS}
The total magnification of a binary-source event is the superposition of two point-lens events, 
\begin{equation}
   A_\lambda = \frac{A_1f_{1,\lambda}+A_2f_{2,\lambda}}{f_{1,\lambda}+f_{2,\lambda}} = \frac{A_{1}+q_{f,\lambda}A_{2}}{1+q_{f,\lambda}},
\end{equation}

\begin{equation}
    q_{f,\lambda} = \frac{f_{2,\lambda}}{f_{1,\lambda}},
\end{equation}
where $f_{{\rm i},\lambda}$ (${\rm i} = 1, 2$) is the flux at wavelength $\lambda$ of each source and $A_\lambda$ is total magnification. We search for 1L2S solutions using MCMC, and the best-fit model is disfavored by $\Delta\chi^2 \sim 38$ compared to the binary-lens ``Wide'' model (see Table \ref{parm3}). Figure \ref{com} presents their cumulative distribution of $\chi^2$ differences, which shows the $\chi^2$ differences are mainly from $\pm20$ days from the peak, rather than outliers. We also consider the microlens-parallax effect, but the improvement is very minor with $\Delta\chi^2 \sim 1.4$. Thus, we exclude the 1L2S solution.



%% file: source.tex
We conduct a Bayesian analysis in Section \ref{lens} to estimate the physical parameters of the lens systems, which requires the constraints of the source properties. Thus, we estimate the angular radius $\theta_*$ and the proper motion of the source in this section.

\subsection{Color-Magnitude Diagram}\label{CMD}
To further estimate the angular Einstein radius $\thetae = \theta_*/\rho$, we estimate the angular radius $\theta_*$ of the source by locating the source on a CMD \citep{Yoo2004}. We calibrate the KMTC02 pyDIA reduction to the OGLE-III star catalog \citep{OGLEIII} and construct a $V - I$ versus $I$ CMD using stars within a $2' \times 2'$ square centered on the event (see Figure \ref{cmd}). The red giant clump is at $(V - I, I)_{\rm cl} = (2.64 \pm 0.01,  16.30\pm 0.02)$, whereas the source is at $(V - I, I)_{\rm S} = (2.40 \pm 0.07, 22.12 \pm 0.05)$ for the Wide solution and $(V - I, I)_{\rm S} = (2.40 \pm 0.07, 22.01 \pm 0.05)$ for the Close solution. We adopt the intrinsic color and de-reddened magnitude of the red giant clump $(V - I, I)_{\rm cl,0} = (1.06, 14.42)$ from \cite{Bensby2013} and \cite{Nataf2016}, and then we derive the intrinsic color and de-reddened brightness of the source as $(V - I, I)_{\rm S,0} = (0.82 \pm 0.08, 20.24 \pm 0.06)$ for the Wide solution and $(V - I, I)_{\rm S,0} = (0.82 \pm 0.08, 20.13 \pm 0.06)$ for the close solution. These values suggest the source is either a late-G or early-K type main-sequence star. Using the color/surface-brightness relation for dwarfs and sub-giants of \cite{Adams2018}, we obtain 

\begin{numcases}{\theta_* =}
0.32 \pm 0.03 ~\mu {\rm as}~{\rm for~the~Wide~solution}, \\
0.34 \pm 0.03 ~\mu {\rm as}~{\rm for~the~Close~solution}.      
\end{numcases}


\subsection{Source Proper Motion}\label{pm}
For \event, the microlens source is too faint to measure its proper motion either from \emph{Gaia} \citep[e.g.,][]{OB171186} or from ground-based data \citep[e.g.,][]{OB170896}. However, we can still estimate the source proper motion by the proper-motion distribution of ``bulge'' stars in the \emph{Gaia} DR2 catalog \citep{Gaia2016AA,Gaia2018AA}. We examine a \emph{Gaia} CMD using the stars within 1 arcmin and derive the proper motion (in the Sun frame) of red giant branch stars ($G < 18.6; B_p - R_p > 2.2$). We remove one outlier and obtain (in the Sun frame) 
\begin{equation}
\langle\bm{\mu}_{\rm bulge}(\ell,b)\rangle = (-6.0, -0.2) \pm (0.2, 0.2)~\text{mas yr}^{-1},
\end{equation}

\begin{equation}
\sigma(\bm{\mu}_{\rm bulge}) = (3.5, 3.0) \pm (0.2, 0.1) ~\text{mas yr}^{-1}.    
\end{equation}

%% file: lens.tex
\subsection{Bayesian Analysis}

For a lensing object, the total mass is related to $\thetae$ and $\pie$ by \citep{Gould1992, Gould2000}
\begin{equation}
    M_{\rm L} = \frac{\thetae}{{\kappa}\pie},
    \label{eq:mass}
\end{equation}
and its distance by
\begin{equation}
    D_{\rm L} = \frac{\mathrm{AU}}{\pie\thetae + \pi_{\rm S}},
\end{equation}
where $\kappa \equiv 4G/(c^2\mathrm{AU}) = 8.144$ mas$/M_{\odot}$, $\pi_{\rm S} = \mathrm{AU}/D_{\rm S}$ is the source parallax, and $D_{\rm S}$ is the source distance. In the present case, neither $\thetae$ nor $\pie$ is unambiguously measured, so we conduct a Bayesian analysis to estimate the physical parameters of the lens systems.

For each solution of $C_\pm$ and $W_\pm$, we first create a sample of $10^9$ simulated events from the Galactic model of \cite{Zhu2017spitzer}. We also choose the initial mass function of \cite{Kroupa2001} and $1.3 M_{\odot}$ for the upper end of the initial mass function. The only exception is that we draw the source proper motions from a Gaussian distribution with the parameters that were derived in Section \ref{pm}. For each simulated event $i$ of solution $k$, we then weight it by
\begin{equation}
    \omega_{{\rm Gal},i,k} = \Gamma_{i,k} \mathcal{L}_{i,k}(\te) \mathcal{L}_{i,k}(\bm{\pi}_{\rm E}) \mathcal{L}_{i,k}(\thetae),
\end{equation}
where $\Gamma_{i,k}\varpropto\theta_{{\rm E},i,k}\times\mu_{{\rm rel},i,k}$ is the microlensing event rate, $\mathcal{L}_{i,k}(\te), \mathcal{L}_{i,k}(\bm{\pi}_{\rm E})$ are the likelihood of its inferred parameters $(\te, \bm{\pi}_{\rm E})_{i,k}$ given the error distributions of these quantities derived from the MCMC for that solution
\begin{equation}
    \mathcal{L}_{i,k}(\te) = \frac{{\rm exp}[-(t_{{\rm E},i,k} - t_{{\rm E},k})^2/2\sigma^2_{t_{{\rm E},k}}]}{\sqrt{2\pi}\sigma_{t_{{\rm E},k}}},
\end{equation}
\begin{equation}
    \mathcal{L}_{i,k}(\bm{\pi}_{\rm E}) = \frac{{\rm exp}[-\sum_{m,n=1}^2b_{m,n}^k(\pi_{{\rm E},m,i}-\pi_{{\rm E},m,k})(\pi_{{\rm E},n,i}-\pi_{{\rm E},n,k})/2]}{2\pi/\sqrt{{\rm det}~b^k}},
\end{equation}
$b_{m,n}^k$ is the inverse covariance matrix of $\bm{\pi}_{{\rm E},k}$, and $(m,n)$ are dummy variables ranging over ($N, E$), and $\mathcal{L}_{i,k}(\thetae)$ is the likelihood derived from the minimum $\chi^2$ for the lower envelope of the ($\chi^2$ vs.~$\rho$) diagram from MCMC and the measured source angular radius $\theta_*$ from Section \ref{CMD}. Finally, we weight each solution by ${\rm exp}(-\Delta\chi^2_k/2)$,  where $\Delta\chi^2_k$ is the $\chi^2$ difference between the $k$th solution and the best-fit solution. 

Table \ref{phy} shows the resulting lens properties and relative weights for each solution, and the combined results. We find that the ``Wide'' solutions are significantly favored because they are preferred by a factor of $\sim {\rm exp}(14/2) \sim 10^3$ from the $\chi^2$ weight, while the ``Wide'' solutions also have slightly higher Galactic model likelihood. The net effect is that the resulting combined solution is basically the same as the wide solution. The Bayesian analysis yields a host mass of $M_{\rm host} = 0.49_{-0.25}^{+0.38}~M_\odot$, a planet mass of $M_{\rm planet} = 2.2_{-1.1}^{+1.9}~M_{J}$, and a host-planet projected separation $r_\perp = 3.5_{-0.9}^{+1.1}~{\rm AU}$, which indicates the planet is a super-Jovian planet well beyond the snow line of an M/K dwarf star (assuming a snow line radius $r_{\rm SL} = 2.7(M/M_{\odot})$~{\rm AU}, \citealt{snowline}). For each solution, the resulting distributions of the lens host-mass $M_{\rm host}$ and the lens distance $D_{\rm L}$ are shown in Figures \ref{baye1} and \ref{baye2}, respectively. The resulting combined distributions of the lens properties are shown in Figure \ref{baye3}.

\subsection{Blended Light}

The light curve analysis shows the blended light for the pySIS light curve is $I_{\rm B} \sim 18.25$. To investigate the blend, we check the higher-resolution $i$-band images (pixel scale $0.185''$, FWHM $\sim 0.6''$) taken from the Canada-France-Hawaii Telescope (CFHT) located at the Maunakea Observatories in 2018 \citep{CFHT}. We identify the source position in the CFHT images from an astrometric transformation of the highly magnified KMTC02 images. We use DoPhot \citep{dophot} to identify nearby stars and do photometry. As a result, DoPhot identifies two stars within $1''$ (see Figure \ref{CFHT}): an $I = 18.18 \pm 0.02$ star offset from the source by $0.88''$, and an $I = 19.43 \pm 0.05$ star offset by $0.61''$. Thus, the blend of pySIS light curve is from unrelated ambient stars. In addition, the total brightness of the source and the lens is fainter than the nearby $I = 19.43 \pm 0.05$ star.

From the CMD analysis and the Bayesian analysis, the source is a late-G or early-K dwarf and the lens is probably an M/K dwarf. Thus, the lens and source may have approximately equal brightness in the near-infrared, therefore follow-up adaptive-optics (AO) observations can potentially strongly constrain the lens brightness and thus the mass and distance of the planetary system \citep{Batista2015,Dave2015,Bhattacharya2018}. In addition, our Bayesian analysis shows that the lens-source relative proper motion is $\mu_{\rm rel} = 3.3_{-0.9}^{+1.5} {\rm mas\,yr^{-1}}$, so the lens and source will be separated by about 40 mas by 2028. Thus, the source and lens can be resolved by the first AO light on next-generation (30 m) telescopes, which have a resolution $\theta \sim 14(D/30{\rm m})^{-1}$ mas in $H$ band.


%% file: dis.tex
We have reported the discovery and analysis of the microlens planet KMT-2016-BLG-1836Lb, for which the $\sim 1$~day, $q \sim 0.004$ planetary perturbation was detected and characterized by KMTNet's $\Gamma \sim 4~{\rm hr}^{-1}$ observations. Many previous works have explored the mass ratio distribution of microlens planets. Of particular note is the work of \cite{Suzuki2016} which discovered a break in the mass-ratio function of planets at $\log q \sim -4$. In addition \cite{OB160596} tested whether observation strategy (survey vs. survey + followup) could affect the observed mass ratio distribution. A full analysis of the mass-ratio distribution for KMTNet planets is well beyond the scope of this work. However, we construct an initial distribution to emphasize the need for such a detailed analysis in the future.

We conduct our analysis on published KMTNet planets discovered in the 2016--2018 seasons and also on the 2016 season alone, since the 2016 season is the most likely to be complete, i.e. have the least publication bias. Including KMT-2016-BLG-1836Lb, there are 13 published microlens planets with KMTNet data from 2016 and 31 published planets from 2016--2018, most of which (19/31 for all the planets from 2016--2018, and 8/13 for planets from 2016) are located in KMTNet's $\Gamma > 1~{\rm hr}^{-1}$ fields\footnote{Actually, only OGLE-2018-BLG-0596Lb was observed at a cadence of $\Gamma \sim 2~{\rm hr}^{-1}$, while other planets were observed at cadences of $\Gamma \geq 4~{\rm hr}^{-1}$.}. The upper and lower panels of Figure \ref{mrall} show the cumulative distributions of planets by log mass ratio $\log q$ for 31 planets from 2016--2018 and 13 planets from 2016, respectively. For each panel, we also show the cumulative distributions of $\log q$ for planets observed at cadences of $\Gamma > 1~{\rm hr}^{-1}$ and $\Gamma \leq 1~{\rm hr}^{-1}$. For events with n degenerate solutions, each solutions are included at a weight of 1/n.

The KMTNet planet sample appears to have a ``mass ratio desert'' at $-3.7 \lesssim \log q \lesssim -3.0$. The only planet ($\log q \sim -3.2$) that appears in this desert is one of the two degenerate solutions for OGLE-2017-BLG-0373Lb. This potential ``mass ratio desert'' cannot be caused by the detection efficiency of KMTNet because eight planets with $\log q < -3.7$ have been detected by $\Gamma > 1~{\rm hr}^{-1}$. However, the sample of planets from \cite{Suzuki2016}, which was subject to a rigorous analysis, does not show any evidence for a mass ratio desert in this range. Likewise, \cite{OB160596} found that the cumulative distributions of $\log q$ are nearly uniformly distributed in $-4.3 < \log q < -2.0$ (i.e., constant number of detections in each bin of equal $\log q$) for a sample including 44 published microlensing planets before 2016 plus OGLE-2016-BLG-0596Lb.

The most likely source of this discrepancy is incompleteness due to publication bias. For example, the number of planets with $\log q~(<-3.7, >-3.7)$ are (2, 11) in 2016, (4, 5) in 2017, and (3, 6) in 2018, which suggests that there are likely be some unpublished planets with $\log q > -3.7$ from 2017 and 2018. This publication bias could result in the missing planets at $-3.7 \lesssim \log q \lesssim -3.0$ and thus the apparent ``mass ratio desert''. 


The core accretion runaway growth scenario predicts that the planets in the mass range 30--100$M_{\earth}$ are rare \citep{Ida2004}. For the typical microlensing lens mass $M_{\rm host} \sim$ 0.3--0.5 $M_{\odot}$, 30--100$M_{\earth}$ corresponds to mass ratio $-3.7 \lesssim \log q \lesssim -3.0$. Thus, the mass ratio distribution from microlensing can be used to test predictions of core accretion theory. \cite{Suzuki2018} found that the MOA mass-ratio distribution from \cite{Suzuki2016} is inconsistent with those predictions. KMTNet enables an independent measurement of this mass ratio distribution. If the potential ``mass ratio desert'' of the KMTNet planet sample is real, it could be consistent with the core accretion theory of planet formation and potentially contradicts \cite{Suzuki2018}. Verifying this apparent ``mass ratio desert'' requires a full statistical analysis of the KMTNet data including detection efficiency and selection biases.

\software{pySIS \citep{pysis}, pyDIA (doi:10.5281/zenodo.268049), emcee \citep{emcee}, DoPhot \citep{dophot}}

\acknowledgments
This research has made use of the KMTNet system operated by the Korea Astronomy and Space Science Institute (KASI) and the data were obtained at three host sites of CTIO in Chile, SAAO in South Africa, and SSO in Australia. This research uses data obtained through the Telescope Access Program (TAP), which has been funded by the National Astronomical Observatories of China, the Chinese Academy of Sciences, and the Special Fund for Astronomy from the Ministry of Finance. H.Y., X.Z., W.Z., W.T. and S.M. acknowledge support by the National Science Foundation of China (Grant No. 11821303 and 11761131004). Work by AG was supported by AST-1516842 and by JPL grant 1500811. AG received support from the European Research Council under the European Unions Seventh Framework Programme (FP 7) ERC Grant Agreement n. [321035]. Work by CH was supported by the grant (2017R1A4A1015178) of National Research Foundation of Korea. Work by P.F. and W.Z. was supported by Canada-France-Hawaii Telescope (CFHT). MTP was supported by NASA grants NNX14AF63G and NNG16PJ32C, as well as the Thomas Jefferson Chair for Discovery and Space Exploration. Wei Zhu was supported by the Beatrice and Vincent Tremaine Fellowship at CITA. Partly based on observations obtained with MegaPrime/MegaCam, a joint project of CFHT and CEA/DAPNIA, at the Canada-France-Hawaii Telescope (CFHT) which is operated by the National Research Council (NRC) of Canada, the Institut National des Science de lUnivers of the Centre National de la Recherche Scientifique (CNRS) of France, and the University of Hawaii.

%% file: table.tex

\begin{table}[htb]
    \centering
    \caption{Best-fit parameters and their $68\%$ uncertainty range from MCMC for four distinct minima shown in Figure \ref{grid}}
    \begin{tabular}{c c c c c}
    \hline
    \hline
    Solutions  & A & B & C & D \\
    \hline
    $t_0$ (${\rm HJD}^{\prime}$) & 7487.58(4) & 7487.67(4) &  7487.39(4) & 7487.26(5) \\
    $u_0$  & 0.062(5) & 0.055(5) & 0.127(13) & 0.045(3) \\
    $\te$  & 49.9(3.4) & 55.3(3.6) & 30.0(2.1) & 64.8(4.2) \\
    $s$ & 0.90(2) & 1.29(2) & 0.89(1) & 1.02(1) \\
    $q (10^{-3})$  & 3.8(4) & 4.1(5) & 0.055(9) & 5.7(9) \\
    $\alpha$ (deg) & 333.1(0.5) & 333.3(0.5) & 150.3(0.7) & 266.3(0.8) \\
    $\rho(10^{-3})$ & $<2.0$ & $<2.8$  & 0.8(2) & 0.4(1) \\
    $I_{\rm S}$ & 22.01(5) & 22.12(5)  & 21.46(4) & 22.28(5) \\
    $I_{\rm B}$ & 18.25(1) & 18.25(1)  & 18.26(1) &  18.25(1) \\
    $\chi^2/dof$ & 9174.1/9154 & 9158.0/9154 & 9631.7/9154 & 9393.0/9154 \\
    \hline
    \hline
    \end{tabular}
    \label{parm1}
\end{table}

\begin{table}[htb]
    \centering
    \caption{Best-fit parameters and their $68\%$ uncertainty range for binary-lens model with parallax}
    \begin{tabular}{c|c c|c c}
    \hline
    \hline
       & \multicolumn{2}{|c|}{Wide} & \multicolumn{2}{c}{Close} \\
    \hline
    Solutions & $W_{+}$ & $W_{-}$ & $C_{+}$ & $C_{-}$ \\
    \hline
    $t_0$ (${\rm HJD}^{\prime}$) & 7487.69(7) & 7487.68(6) & 7487.60(4) & 7487.61(4) \\
    $u_0$  & 0.053(5) & $-$0.056(4)  & 0.061(5)  & $-$0.061(4) \\
    $\te$  & 56.2(3.9) & 54.2(2.9) & 50.0(3.1) & 49.5(2.7) \\
    $s$  & 1.31(3) & 1.30(2) & 0.89(2) & 0.88(2) \\
    $q (10^{-3})$  & 4.6(9) & 4.5(8) & 4.3(6) & 4.4(6) \\
    $\alpha$ (deg)  & 335.1(2.0) & 25.4(1.7) & 335.1(1.3) & 24.7(1.1) \\
    $\rho(10^{-3})$  & $<2.7$ & $<2.7$ & $<2.2$ & $<2.2$ \\
    $\pi_{\rm E,N}$  & 0.56(0.59) & $-$0.46(0.56) & 0.66(40) & $-$0.79(37) \\
    $\pi_{\rm E,E}$  & 0.08(8) & 0.05(8) & 0.07(10) & 0.02(8) \\
    $I_{\rm S}$  & 22.14(5) & 22.10(4) & 22.01(5) &22.00(4) \\
    $I_{\rm B}$  & 18.25(1) & 18.25(1) & 18.25(1) & 18.25(1) \\
    $\chi^2/dof$ & 9156.8/9152 & 9156.3/9152 & 9171.9/9152 & 9171.1/9152 \\
    \hline
    \hline
    \end{tabular}
    \label{parm2}
\end{table}

\begin{table*}[htb]
    \centering
    \caption{Best-fit parameters and their $68\%$ uncertainty range from MCMC for binary-source models}
    \begin{tabular}{c c c c}
    \hline
    \hline
      & & \multicolumn{2}{c}{Parallax models} \\
    Solution & Standard & $u_0 > 0$ & $u_0 < 0$\\
    \hline
    $t_{0,1}$ (${\rm HJD}^{\prime}$) & 7487.12(2) & 7487.17(4)  & 7487.15(4) \\
    $t_{0,2}$ (${\rm HJD}^{\prime}$) & 7494.73(3) & 7493.78(3) & 7493.77(3)\\
    $u_{0,1}$  & 0.046(2) & 0.048(2)  & $-$0.046(2)\\
    $u_{0,2}$  & 0.002(2) & 0.002(2) & $-$0.002(2)\\
    $t_E ({\rm d})$ & 65.02(3) & 64.98(6)  & 65.02(6) \\
    $\rho_{1}$ & 0.012(10) & 0.017(13)  & 0.014(12)\\
    $\rho_{2}$ & 0.0045(13) & 0.0043(9) & 0.0046(11)\\
    $q_{f,I}$ & 0.038(3) & 0.034(5) & 0.037(4)\\
    $I_{\rm S}$ & 22.36(16) & 22.35(14) & 22.36(14) \\
    $I_{\rm B}$ & 18.25(1) & 18.25(1) & 18.25(1)  \\
    \hline
    $\chi^2/dof$  & 9196.2/9153 &  9194.8/9151 & 9194.2/9151 \\
    \hline
    \end{tabular}\\
    \label{parm3}
\end{table*}

\begin{table}[htb]
    \centering
    \caption{Physical parameters for \event}
    \begin{tabular}{c|c c c c|c c}
    \hline
    \hline
     & \multicolumn{4}{c|}{Physical Properties} & \multicolumn{2}{c}{Relative Weights} \\
    \hline
    Solutions  & $M_{\rm host}[M_{\odot}]$ & $M_{\rm planet}[M_{J}]$ & $D_{\rm L}$[kpc] & $r_{\bot}$[AU] & Gal.Mod. & $\chi^2$ \\
    \hline
    $W_{+}$ & $0.48^{+0.39}_{-0.25}$ & $2.2^{+1.9}_{-1.2}$ & $7.1^{+0.8}_{-2.6}$ & $3.6^{+1.2}_{-0.9}$ & 0.928 & 0.779 \\

    $W_{-}$ & $0.49^{+0.39}_{-0.25}$ & $2.2^{+1.8}_{-1.1}$ & $7.2^{+0.8}_{-2.2}$ & $3.5^{+1.1}_{-0.9}$ & 1.000 & 1.000 \\

    $W_{\rm Total}$ & $0.49^{+0.38}_{-0.25}$ & $2.2^{+1.9}_{-1.1}$ & $7.1^{+0.8}_{-2.4}$ & $3.5^{+1.1}_{-0.9}$ &  & \\

    \hline

    $C_{+}$ & $0.51^{+0.41}_{-0.28}$ & $2.2^{+1.8}_{-1.2}$ & $6.3^{+1.3}_{-2.7}$ & $2.7^{+0.7}_{-0.7}$ & 0.844 & 0.0004 \\

    $C_{-}$ & $0.60^{+0.40}_{-0.32}$ & $2.7^{+1.8}_{-1.4}$ & $7.1^{+0.7}_{-1.8}$ & $2.6^{+0.7}_{-0.6}$ & 0.247 & 0.0006\\

    $C_{\rm Total}$ & $0.53^{+0.42}_{-0.29}$ & $2.3^{+1.9}_{-1.3}$ & $6.5^{+1.2}_{-2.8}$ & $2.7^{+0.7}_{-0.7}$ &  & \\

    \hline

    Total & $0.49^{+0.38}_{-0.25}$ & $2.2^{+1.9}_{-1.1}$ & $7.1^{+0.8}_{-2.4}$ & $3.5^{+1.1}_{-0.9}$ &  & \\
    \hline
    \hline
    \end{tabular}\\
    \label{phy}
\end{table}

%% file: figure.tex
\begin{figure*}[htbp]
    \centering
    \subfigure{
    \begin{minipage}{16.5cm}
    \centering
    \includegraphics[width=\columnwidth]{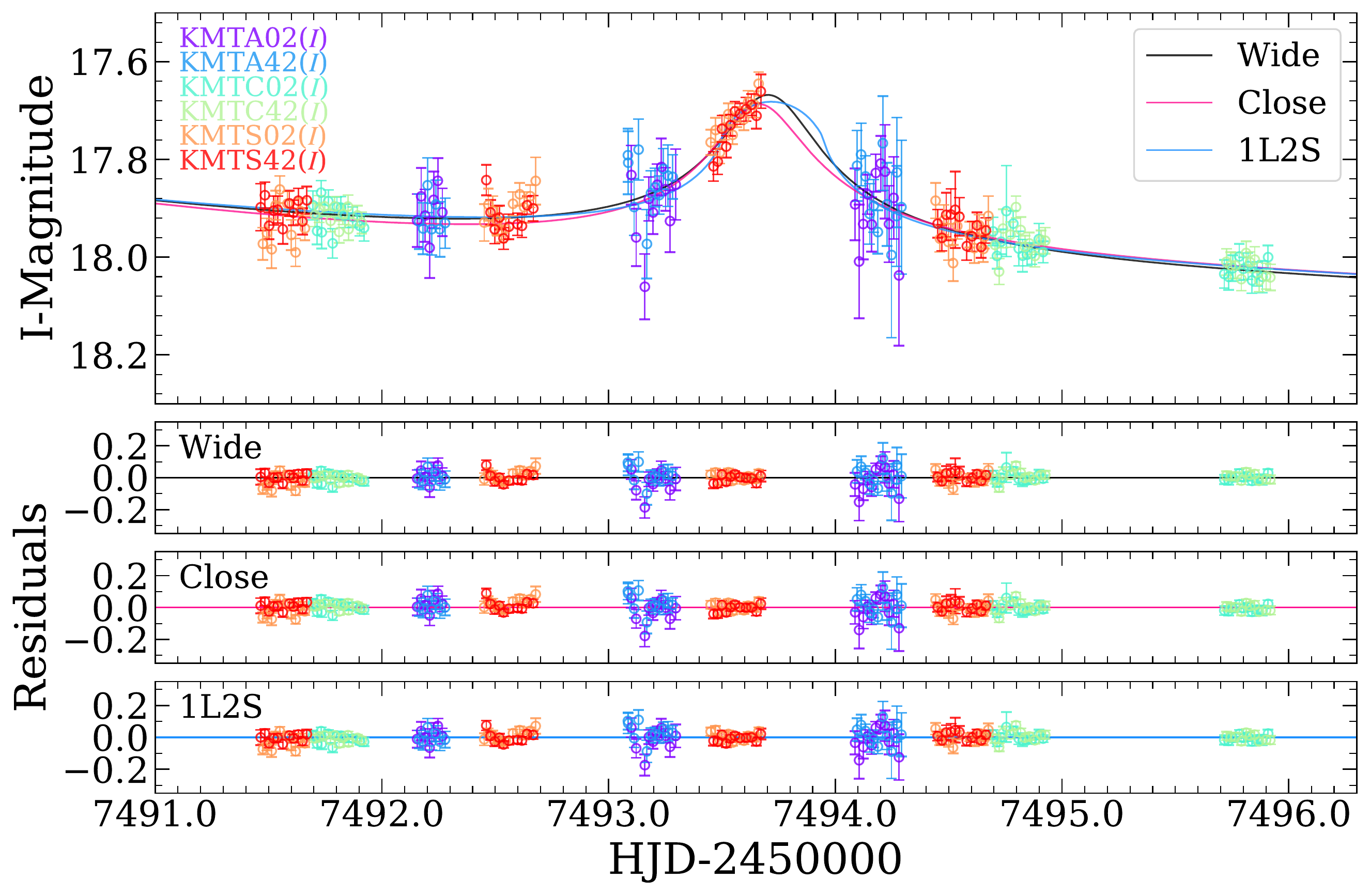}
    \end{minipage}
    }
    
    \subfigure{
    \begin{minipage}{16.5cm}
    \centering
    \includegraphics[width=\columnwidth]{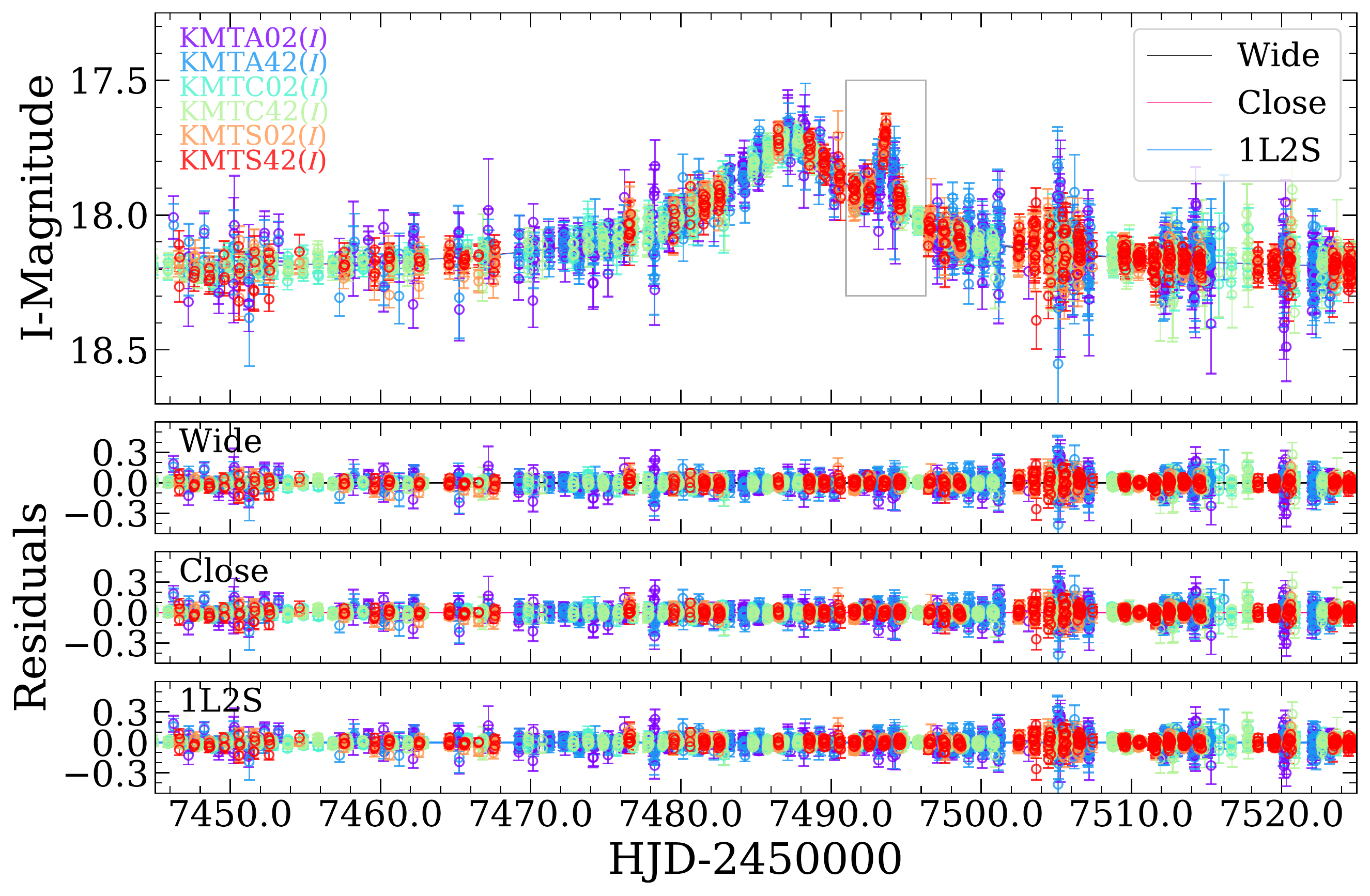}
    \end{minipage}
    }
    \caption{The data of \event\ together with the best-fit models of the binary-lens ``Wide'', binary-lens ``Close'', and binary-source (1L2S) model. The upper panel shows a zoom of the anomaly. The residuals for each model are shown separately. The light curve and data have been calibrated to standard $I$-band magnitude.}
    \label{lc}
\end{figure*}

\begin{figure*}[htb]
    \centering
    \subfigure{
    \begin{minipage}{15cm}
    \centering
    \includegraphics[width=\columnwidth]{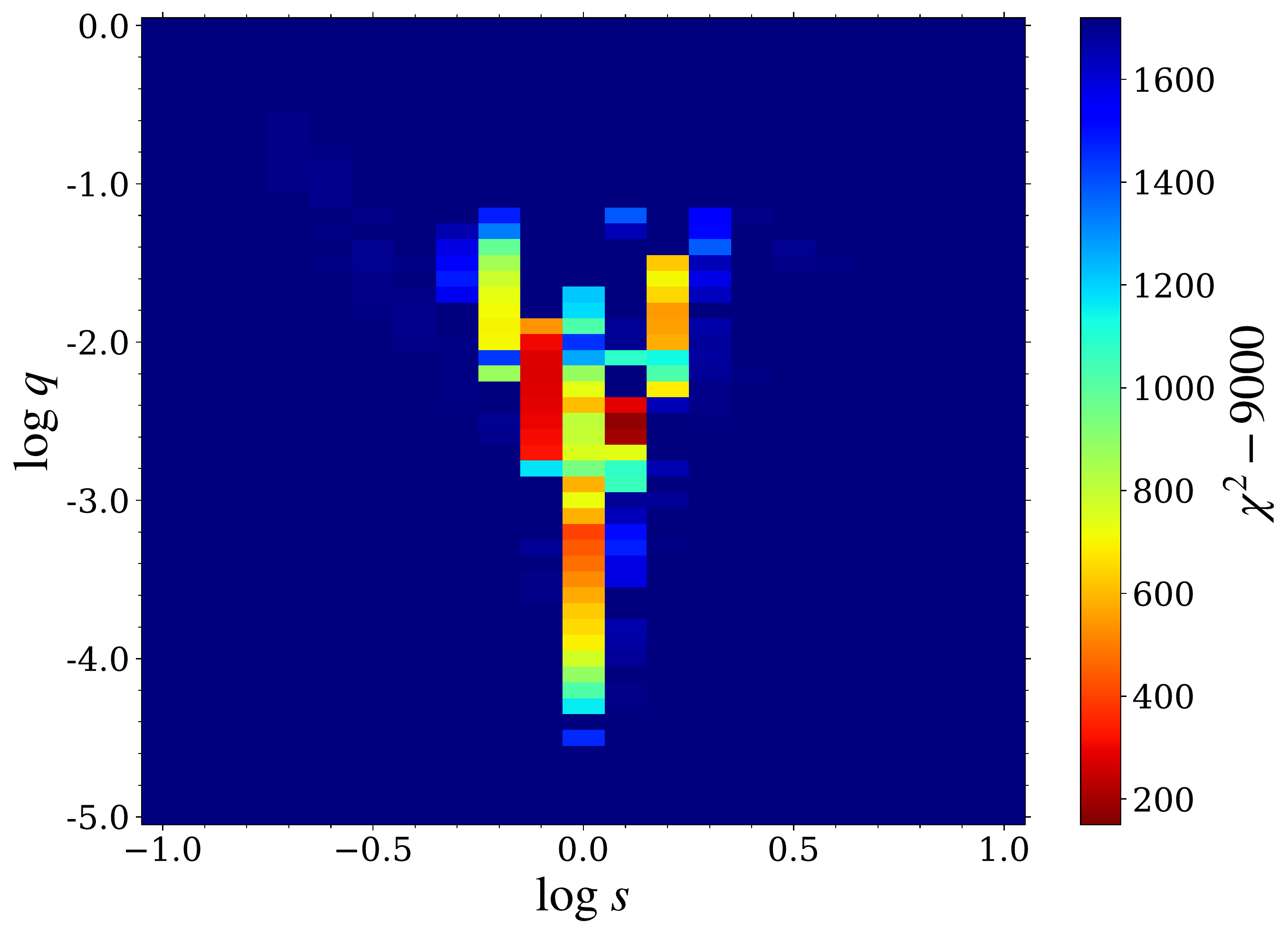}
    \end{minipage}
    }
    \subfigure{
    \begin{minipage}{15cm}
    \centering
    \includegraphics[width=\columnwidth]{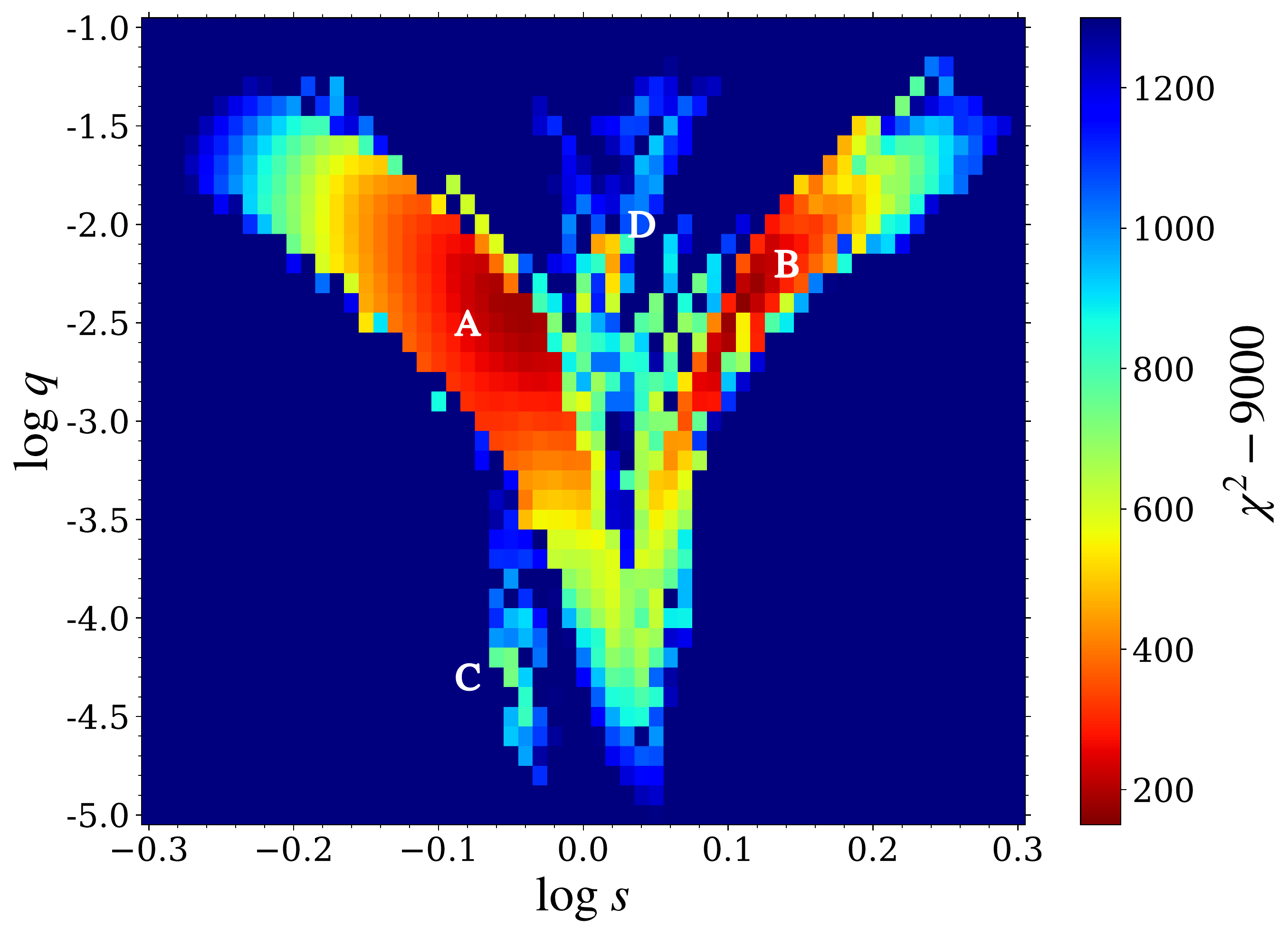}
    \end{minipage}
    }
    \caption{$\chi^2$ surface in the ($\log s, \log q$) plane drawn from the grid search. The upper panel shows the space that is equally divided on a ($21 \times 51$) grid with ranges of $-1.0\leq\log s \leq1.0$ and $-5.0\leq \log q \leq0$, respectively. The lower panel shows the space that is equally divided on a ($61 \times 41$) grid with ranges of $-0.3\leq\log s \leq0.3$ and $-5.0\leq \log q \leq-1.0$, respectively. The labels ``A'', ``B'', ``C'' and ``D'' in the lower panel show four distinct minima.}
    \label{grid}
\end{figure*}

\newpage
\begin{figure*}[htbp]
    \centering
    \subfigure{
    \begin{minipage}{16.5cm}
    \centering
    \includegraphics[width=\columnwidth]{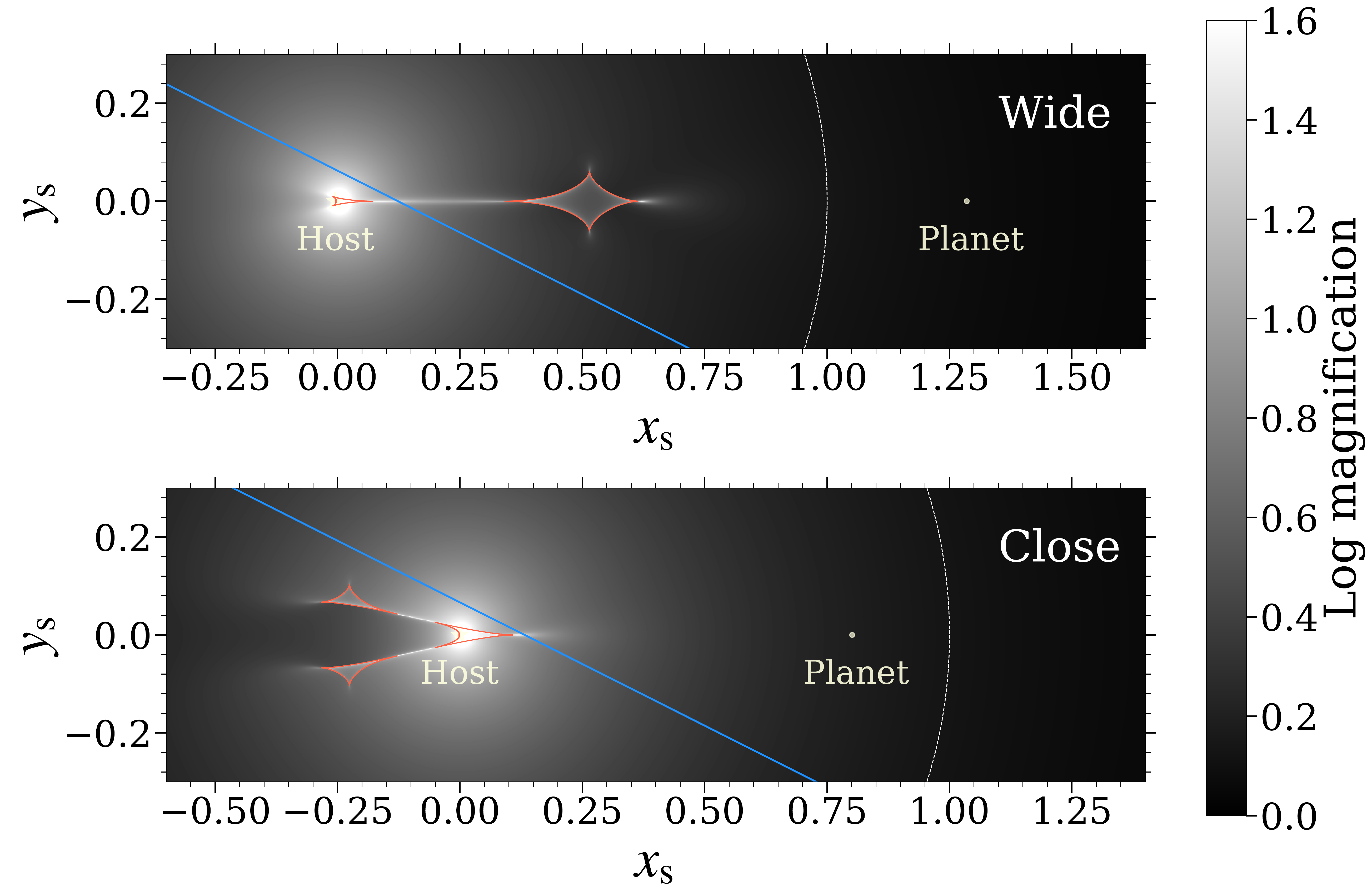}
    \end{minipage}
    }
    \caption{Magnification maps of the standard ``Wide'' (upper panel) and ``Close'' (lower panel) models shown in Table \ref{parm1}. In each panel, the blue line with arrow represents the trajectory of the source with direction. The red contours are the caustics. The dashed lines indicate the Einstein ring and both $x_{S}$ and $y_{S}$ are in unit of the Einstein radius. The grayscale indicates the magnification of a point source at each position, where white means higher magnification.}
    \label{cau}
\end{figure*}

\begin{figure}[htb] 
    \includegraphics[width=\columnwidth]{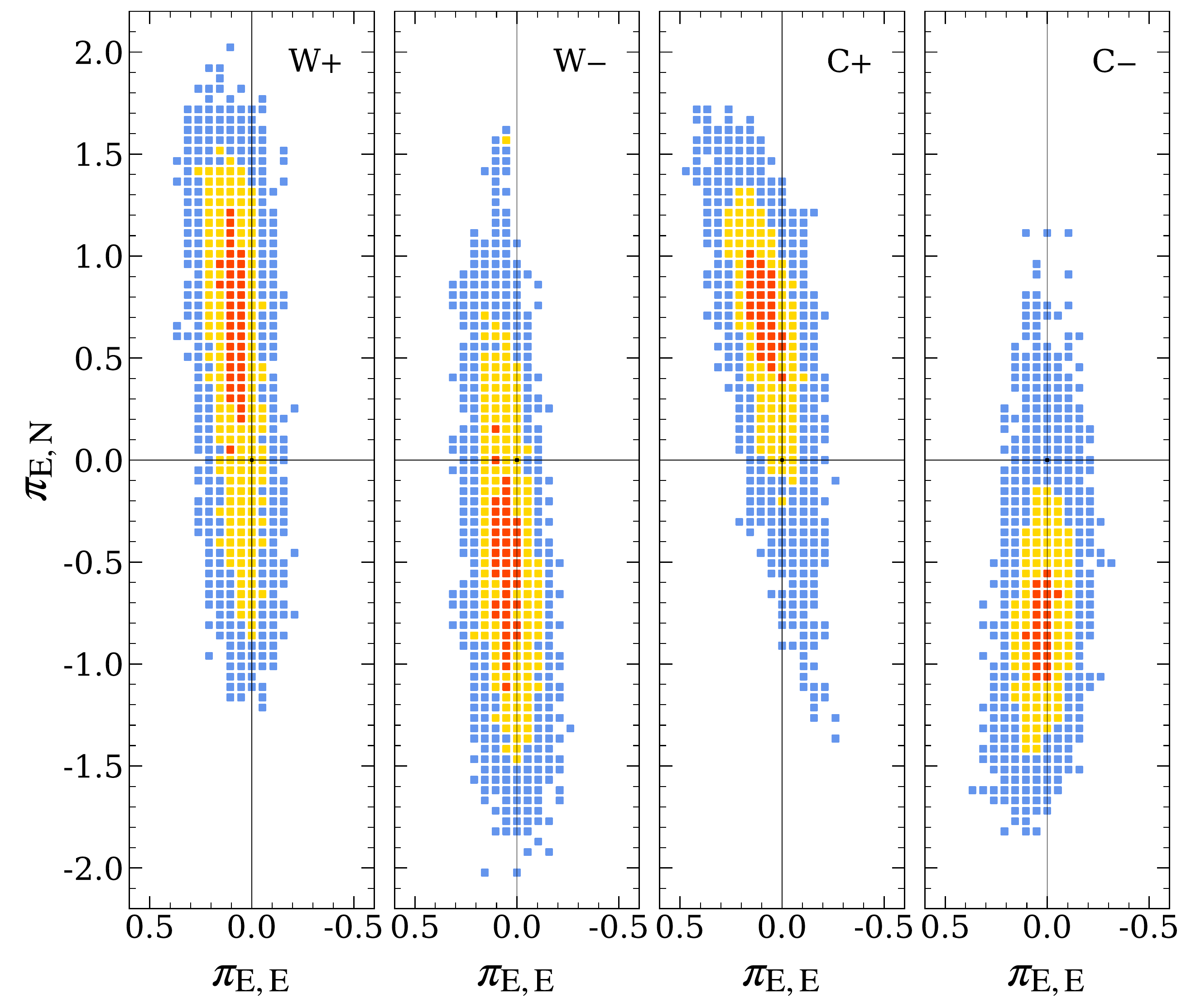}
    \caption{Likelihood distributions for $\bm{\pi_{\rm E}}$ derived from MCMC for $W_{\pm}$ and $C_{\pm}$ solutions (see Table \ref{parm1} for the solution parameters). Red, yellow, and blue show likelihood ratios $[-2\Delta\ln{\mathcal{L}/\mathcal{L}_{\rm max}}] < (1, 4, \infty)$, respectively.}
    \label{pie}
\end{figure}

\newpage
\begin{figure*}[htbp]
    \centering
    \subfigure{
    \begin{minipage}{16.5cm}
    \centering
    \includegraphics[width=\columnwidth]{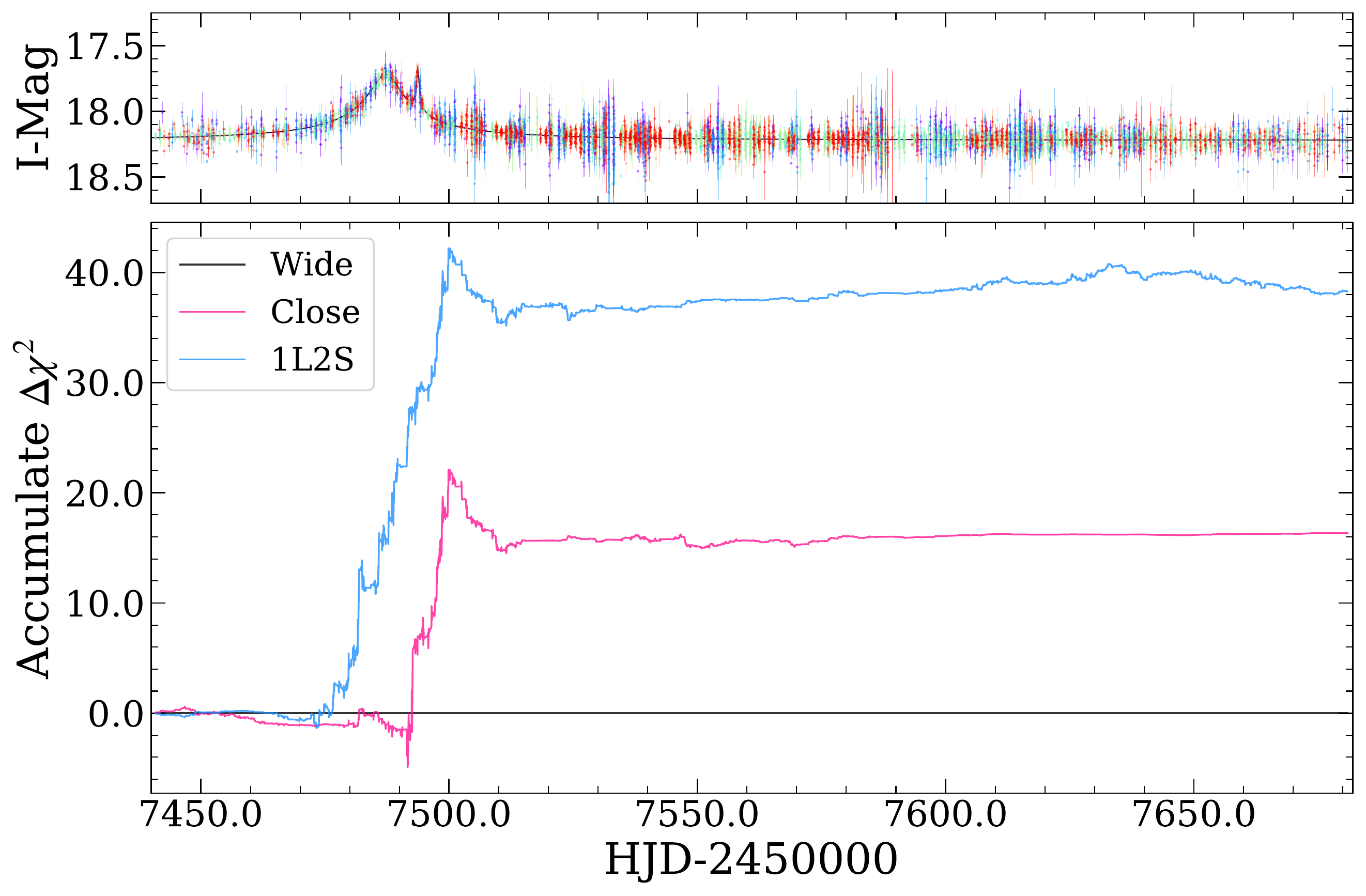}
    \end{minipage}
    }
    \caption{Cumulative distribution of $\chi^2$ differences ($\Delta\chi^2 = \chi^2_{\rm model} - \chi^2_{\rm Wide}$) between the ``Close'', binary-source (1L2S), and the ``Wide'' models.}
    \label{com}
\end{figure*}

\begin{figure}[htb] 
    \includegraphics[width=\columnwidth]{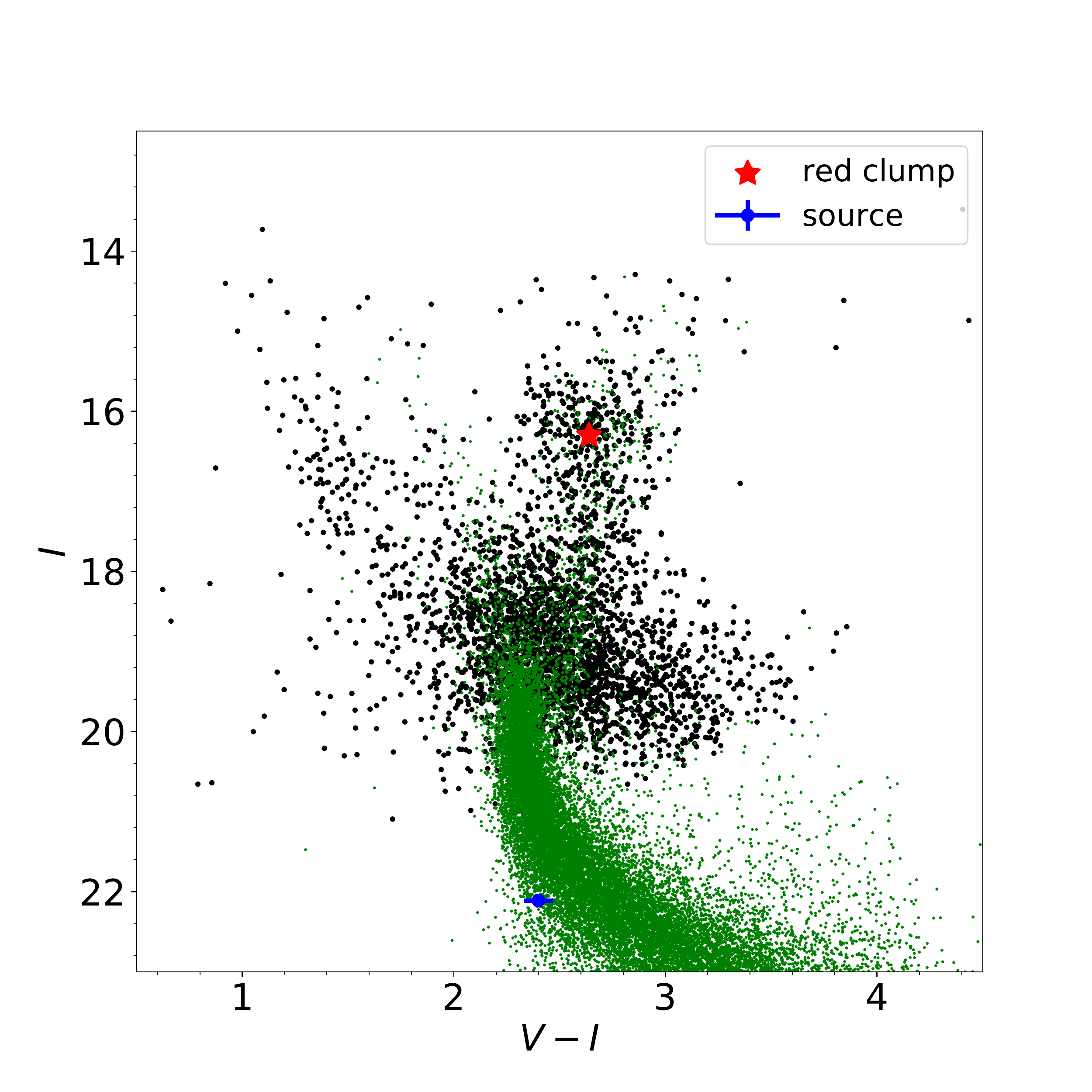}
    \caption{Color-magnitude diagram of a $2′ \times 2′$ square centered on  \event. The black dots show the stars from pyDIA photometry of KMTC02 data which are calibrated to OGLE-III star catalog \citep{OGLEIII}, and the green dots show the HST CMD of \cite{HSTCMD} whose red-clump centroid is adjusted to match pyDIA’s using the Holtzman field red-clump centroid of$(V - I, I)=(1.62, 15.15)$ \citep{MB07192}. The red asterisk shows the centroid of the red clump, and the blue dot indicates the position of the source.}
    \label{cmd}
\end{figure}

\clearpage

\begin{figure}[htb] 
    \includegraphics[width=\columnwidth]{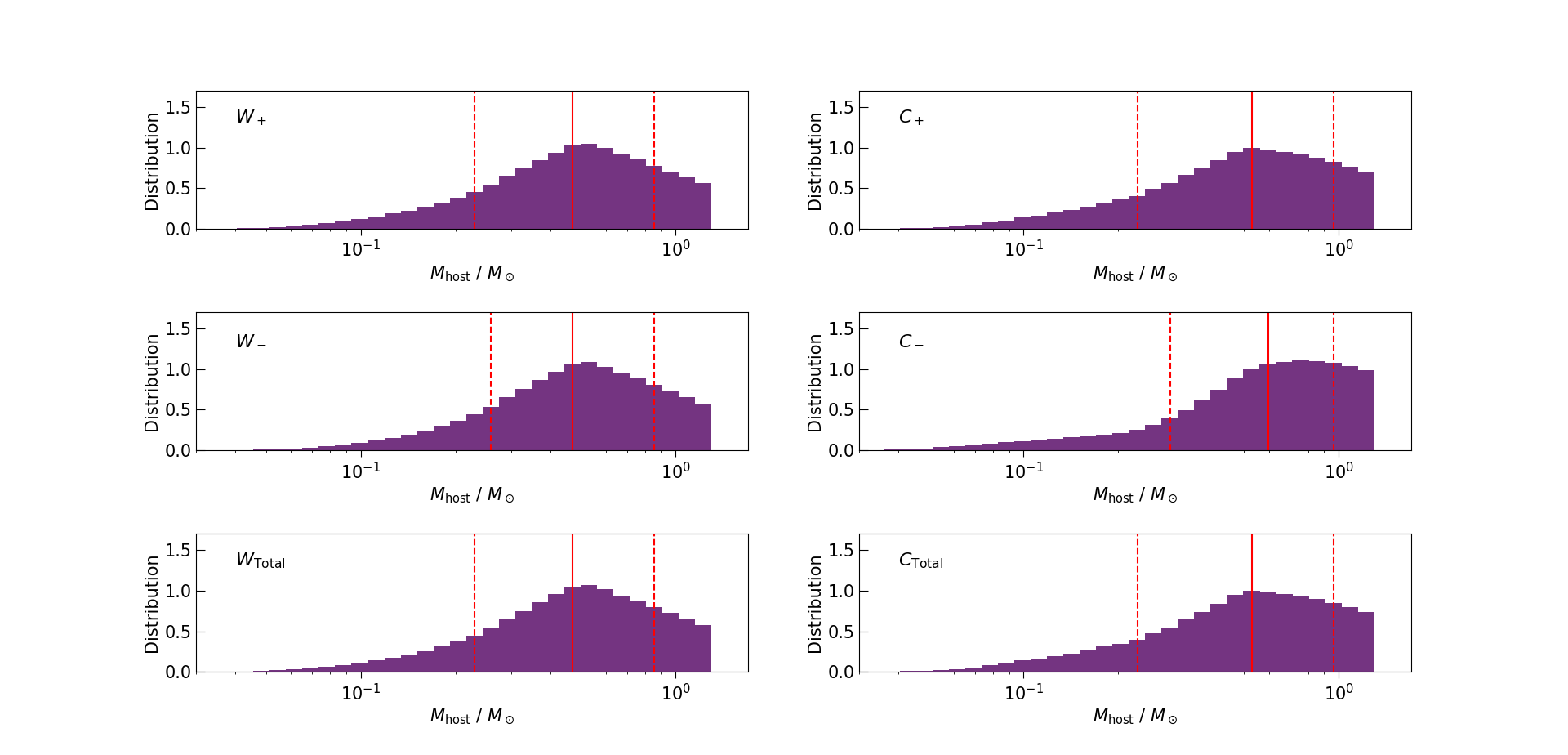}
    \caption{Bayesian posterior distributions of the lens host-mass $M_{\rm host}$ for each solution of $C_\pm$ and $W_\pm$ (top two rows) and the combined distributions for $C_\pm$ and $W_\pm$ (bottom row). In each panel, the red solid vertical line represents the median value and the two red dashed lines represent 16th and 84th percentiles of the distribution.}
    \label{baye1}
\end{figure}

\begin{figure}[htb] 
    \includegraphics[width=\columnwidth]{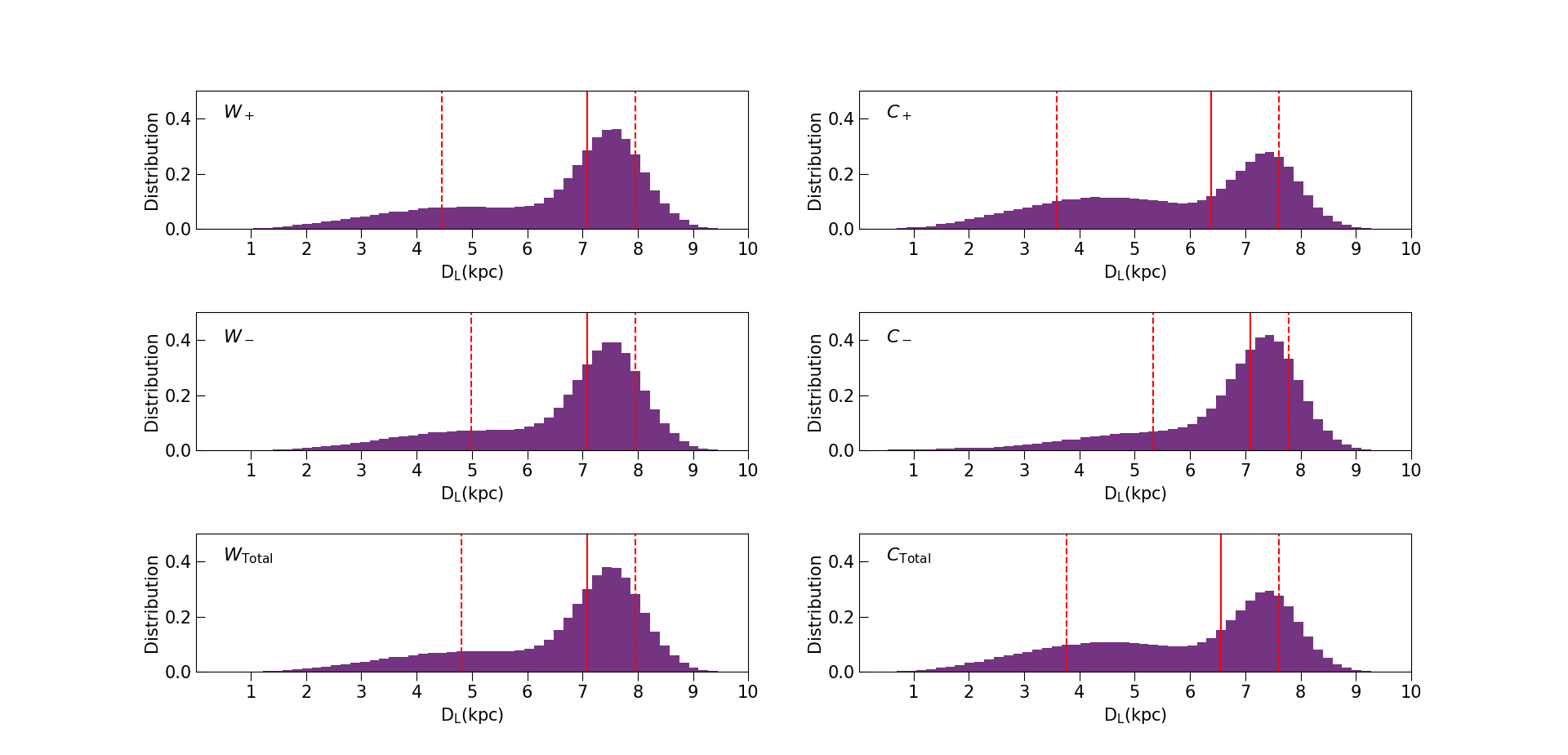}
    \caption{Bayesian posterior distributions of the lens distance $D_{\rm L}$. The plot is similar to Figure \ref{baye1}.}
    \label{baye2}
\end{figure}

\begin{figure*}[htb]
    \centering
    \subfigure{
    \begin{minipage}{11cm}
    \centering
    \includegraphics[width=\columnwidth]{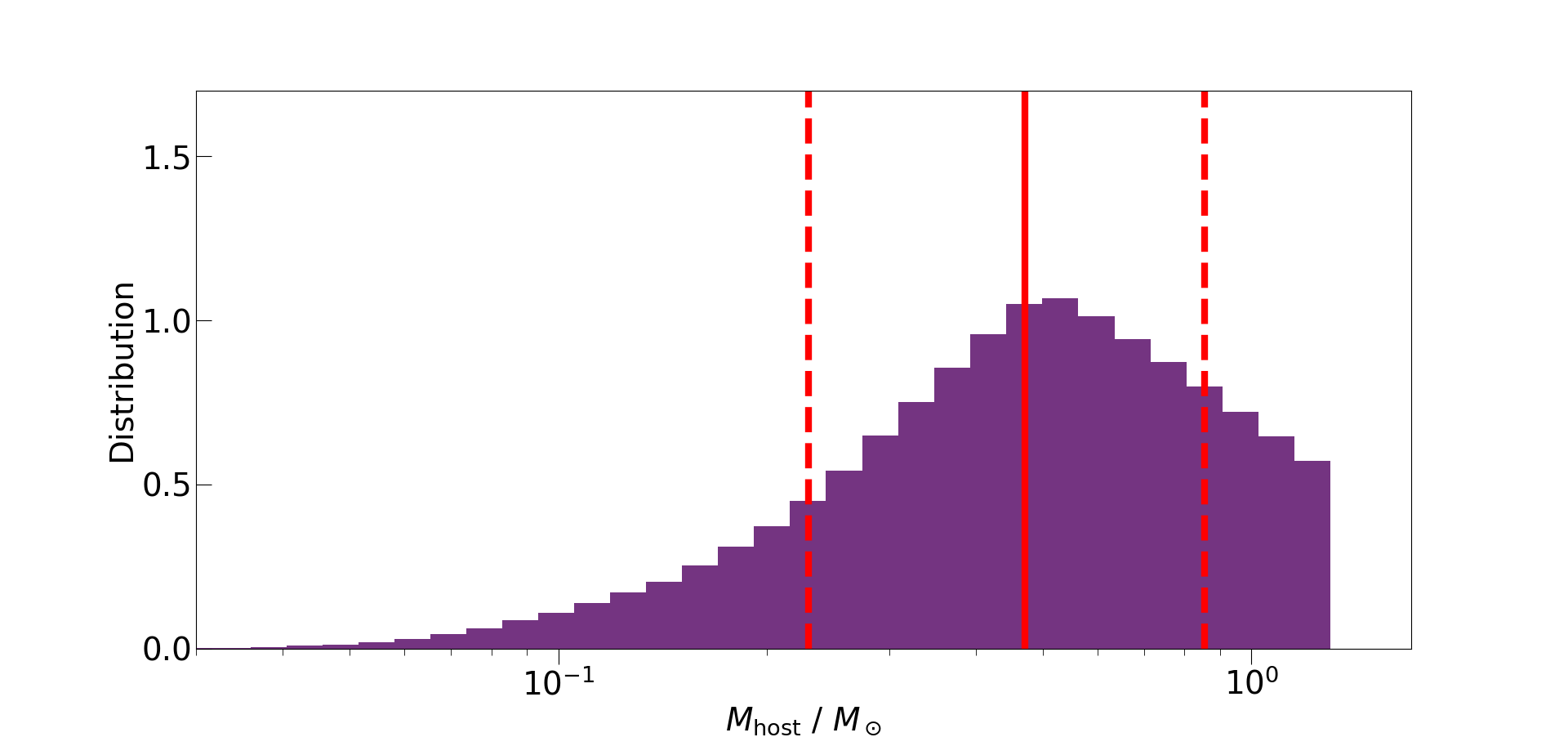}
    \end{minipage}
    }
    \subfigure{
    \begin{minipage}{11cm}
    \centering
    \includegraphics[width=\columnwidth]{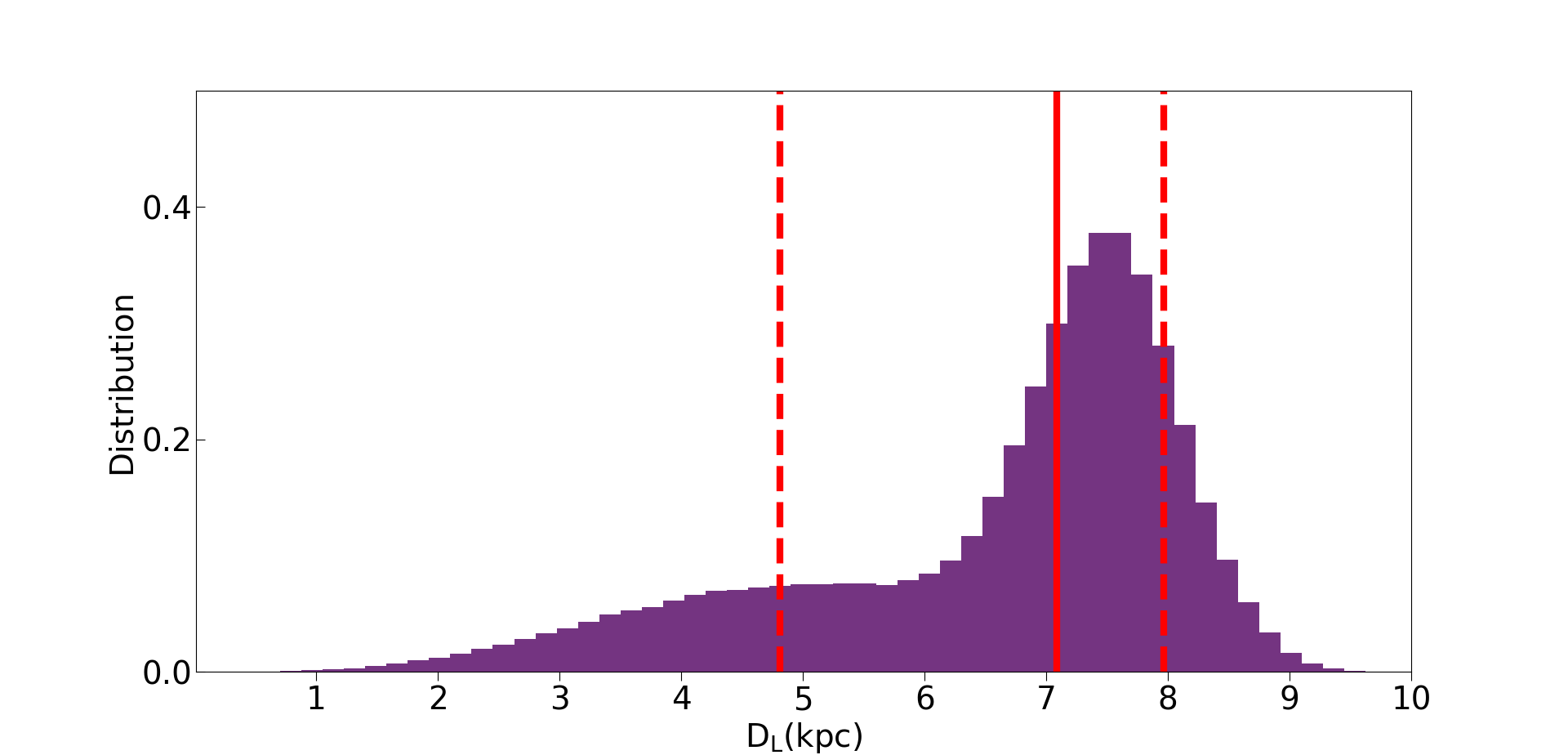}
    \end{minipage}
    }
    \subfigure{
    \begin{minipage}{11cm}
    \centering
    \includegraphics[width=\columnwidth]{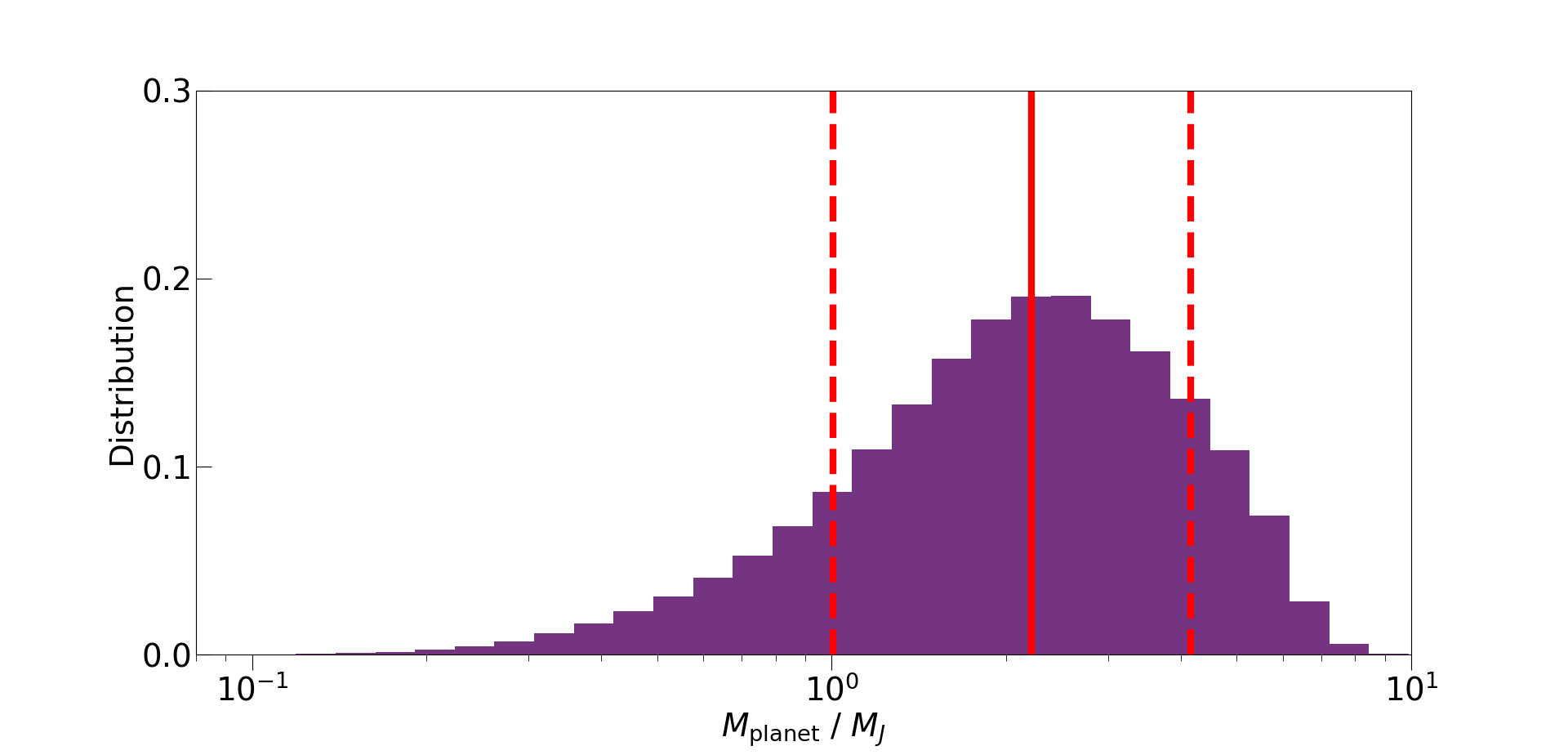}
    \end{minipage}
    }
    \subfigure{
    \begin{minipage}{11cm}
    \centering
    \includegraphics[width=\columnwidth]{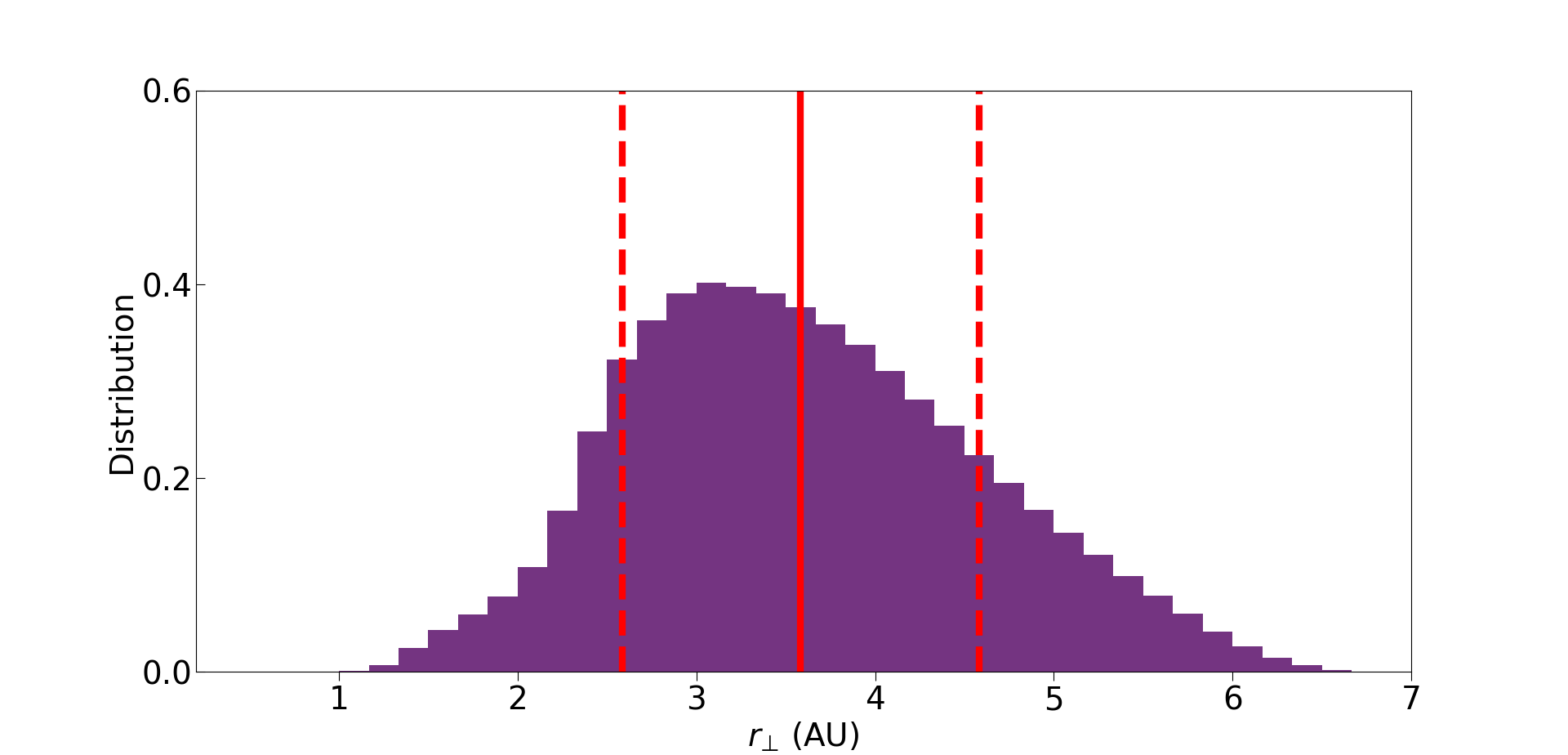}
    \end{minipage}
    }
    \caption{The combined Bayesian distributions of the lens host-mass $M_{\rm host}$, the lens distance $D_{\rm L}$, the planet-mass $M_{\rm planet}$, and the projected separation $r_\perp$ of the planet.}
    \label{baye3}
\end{figure*}

\newpage
\begin{figure*}[htbp]
    \centering
    \subfigure{
    \begin{minipage}{16.5cm}
    \centering
    \includegraphics[width=\columnwidth]{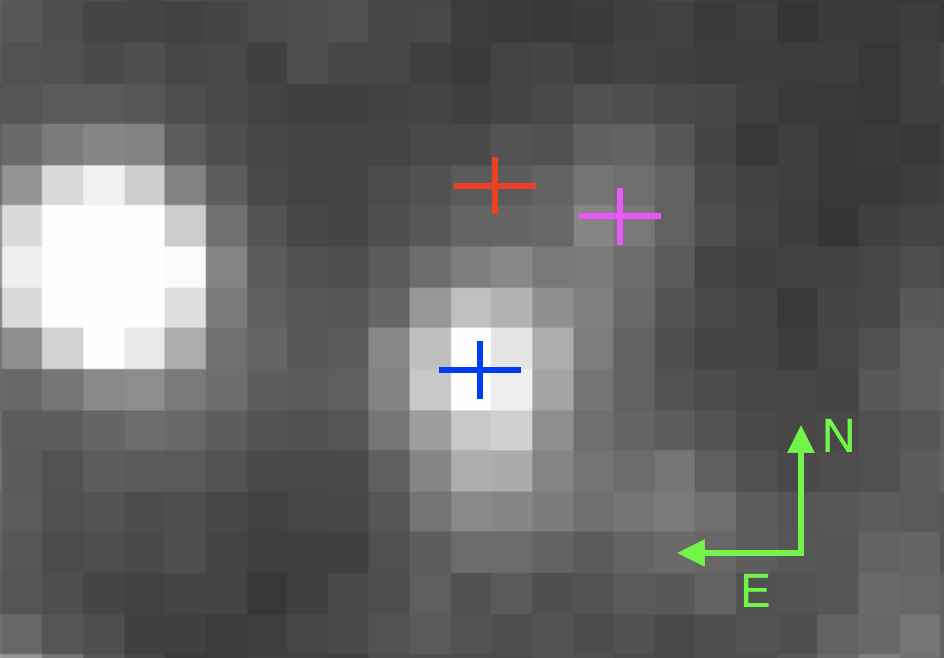}
    \end{minipage}
    }
    \caption{$i$-band CFHT images within $4.9'' \times 3.0''$ around the event. The red cross indicates the source position derived from an astrometric transformation of the highly magnified KMTC02 images. The blue and magenta crosses indicate the $I = 18.18 \pm 0.02$ star and $I = 19.43 \pm 0.05$ star found by DoPhot \citep{dophot}, respectively.}
    \label{CFHT}
\end{figure*}

\begin{figure*}[htb]
    \centering
    \subfigure{
    \begin{minipage}{15cm}
    \centering
    \includegraphics[width=\columnwidth]{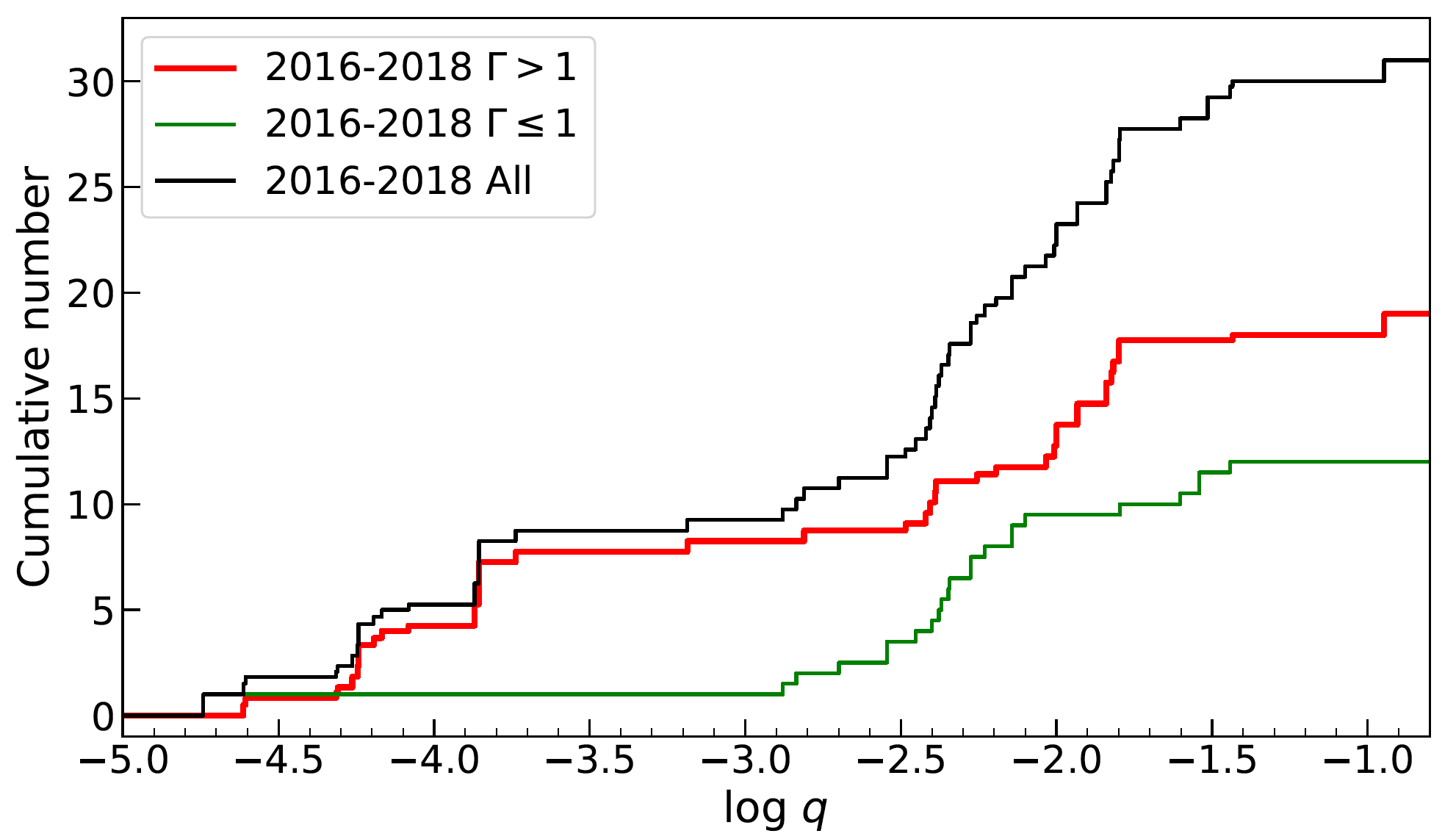}
    \end{minipage}
    }
    \subfigure{
    \begin{minipage}{15cm}
    \centering
    \includegraphics[width=\columnwidth]{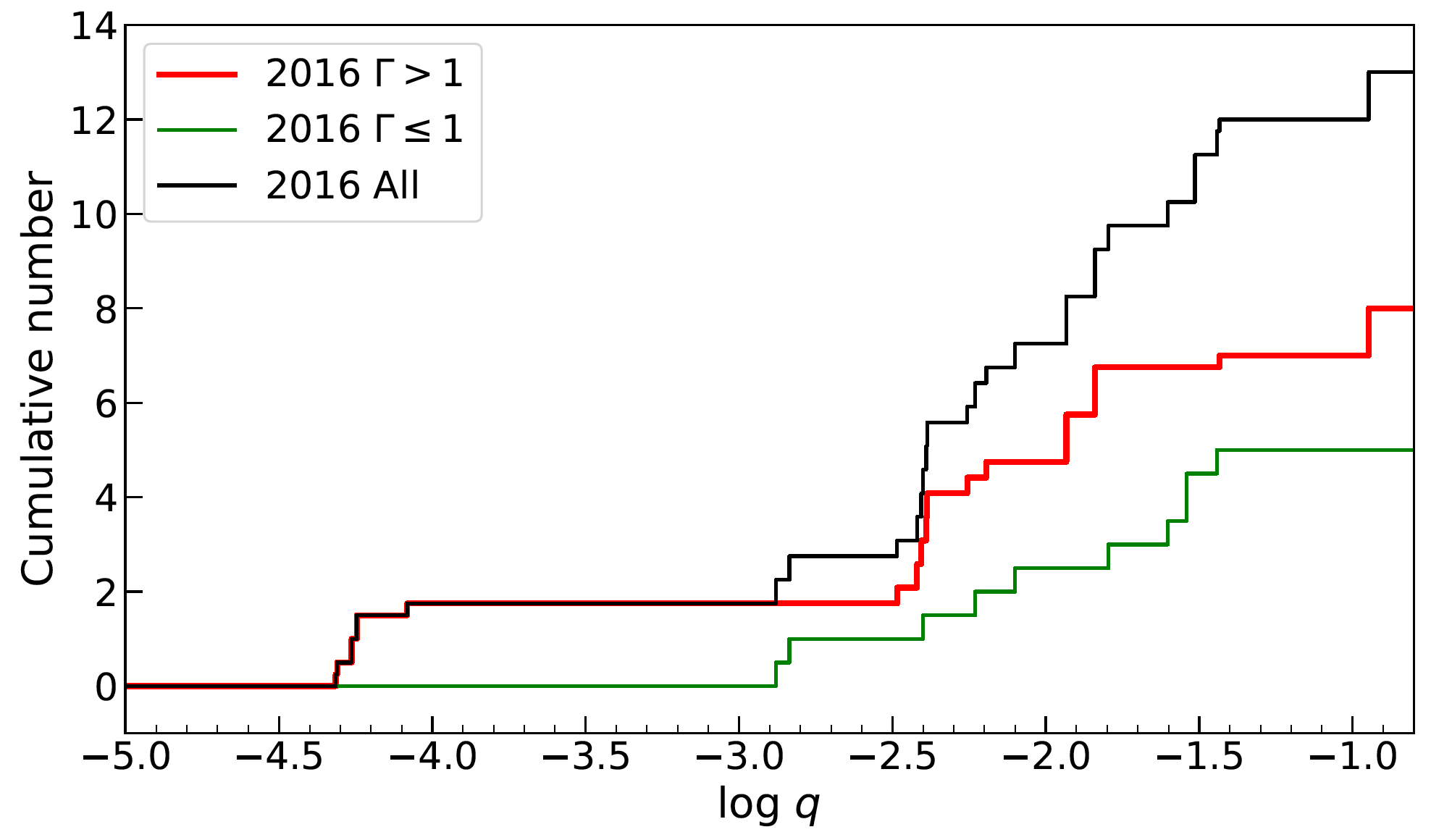}
    \end{minipage}
    }
    \caption{Cumulative distributions of 31 published KMTNet microlensing planets from 2016--2018 by $\log q$ (upper panel) and 13 published KMTNet microlensing planets from 2016 by $\log q$ (lower panel). In each panel, the red and green lines represent the distributions for planets observed at cadences of $\Gamma > 1~{\rm hr}^{-1}$, $\Gamma \leq 1~{\rm hr}^{-1}$, respectively, and the black line represents the distribution of all the planets.}
    \label{mrall}
\end{figure*}